\documentclass[letterpaper,twocolumn,10pt]{article}								
						
\usepackage{usenix}
\usepackage{tikz}
\usepackage{amsmath}

\newcommand{\tool}{\mbox{\textsc{Moat}}\xspace}

\newcommand{\fixme}[1]{{\color{black}{#1}}} 
\newcommand{\mr}[1]{{\color{black}{#1}}}

\newcommand{\parh}[1]{\noindent\textbf{#1}}

\usepackage[numbers,sort&compress]{natbib}
\usepackage{xspace}
\usepackage{amsmath,amssymb,amsfonts}
\usepackage{authblk}
\usepackage{graphicx}
\usepackage{textcomp}
\usepackage{xcolor}
\usepackage{threeparttable}
\usepackage{colortbl}
\usepackage{graphicx}
\usepackage{listings}
\usepackage{color}
\usepackage{array}
\usepackage{float}
\usepackage{graphicx}
\usepackage{caption}
\usepackage{subcaption}
\usepackage{multirow}
\usepackage{colortbl}
\usepackage[linesnumbered,boxed]{algorithm2e}
\usepackage{algpseudocode}
\usepackage{framed}
\usepackage{pifont}
\usepackage{enumitem}
\usepackage{balance}
\usepackage{url}

\usepackage{breakurl}
\usepackage[T1]{fontenc}
\usepackage[utf8]{inputenc}
\usepackage[font=small,labelfont=bf,tableposition=top]{caption}
\usepackage{booktabs}
\usepackage{tabularx}
\usepackage{diagbox}
\usepackage{algpseudocode}
\usepackage{amsmath}
\usepackage{listings}
\usepackage{makecell}
\usepackage{verbatim}
\usepackage{acronym}
\usepackage[super]{nth}
\usepackage{tikz}
\usepackage{amsmath}
\usepackage{filecontents}
\usepackage{pifont}
\usepackage{bold-extra}
\usepackage{tikz,pgf}
\usetikzlibrary{calc}
\usepackage{hyperref}
\makeatletter
\def\mfontsize{\f@size}
\newcommand*\circled[1]{\tikz[baseline=(char.base)]{
		\node[shape=circle,draw,inner sep=0pt] (char) {#1};}}

\newcommand{\F}{Fig.}

\newcommand{\T}{Table}
\renewcommand{\S}{Sec.}

\newcommand{\customfootnotetext}[2]{{
		\renewcommand{\thefootnote}{#1}
		\footnotetext[0]{#2}}}

\usepackage{xcolor}

\definecolor{codegreen}{rgb}{0,0.6,0}
\definecolor{codegray}{rgb}{0.5,0.5,0.5}
\definecolor{codepurple}{rgb}{0.58,0,0.82}
\definecolor{backcolour}{rgb}{0.95,0.95,0.92}

\lstdefinestyle{mystyle}{
	backgroundcolor=\color{backcolour},   
	commentstyle=\color{codegreen},
	keywordstyle=\color{magenta},
	numberstyle=\tiny\color{codegray},
	stringstyle=\color{codepurple},
	frame=lrtb,
	basicstyle=\ttfamily\footnotesize,
	breakatwhitespace=false,         
	breaklines=true,                 
	captionpos=b,                    
	keepspaces=true,                 
	numbers=left,                    
	numbersep=5pt,                  
	showspaces=false,                
	showstringspaces=false,
	showtabs=false,                  
	tabsize=2,
	 xleftmargin=\dimexpr\fboxsep-\fboxrule+5pt,
	 xrightmargin=\dimexpr\fboxsep-\fboxrule
}

\usepackage[normalem]{ulem}

\begin{document}
	
	\date{}
	
	\title{\Large \bf \tool: Towards Safe BPF Kernel Extension}
	
\author{Hongyi Lu$^{1,2,3}$, Shuai Wang$^{3,\dagger}$, Yechang Wu$^{2}$, 
Wanning He$^{2}$, Fengwei Zhang$^{2,1,\dagger}$}
\affil{\small
	$^1$\textit{Research Institute of Trustworthy Autonomous Systems, Southern 
	University of Science and Technology} \\
	$^2$\textit{Department of Computer Science and Engineering, Southern 
	University of Science and Technology} \\
	$^3$\textit{Department of Computer Science and Engineering, Hong Kong 
	University of Science and 
	Technology}
}
	
	\maketitle
	\customfootnotetext{$^{\dagger}$}{Shuai Wang and Fengwei Zhang are the corresponding 
	authors.}
  \begin{abstract}
    The Linux kernel extensively uses the Berkeley Packet Filter (BPF) to allow
    user-written BPF applications to execute in the kernel space. The BPF
    employs a verifier to check the security of user-supplied BPF code
    statically. Recent attacks show that BPF programs can evade security checks
    and gain unauthorized access to kernel memory, indicating that the
    verification process is not flawless.
    In this paper, we present \tool, a system that isolates potentially
    malicious BPF programs using Intel Memory Protection Keys (MPK). Enforcing
    BPF program isolation with MPK is not straightforward; \tool\ is designed
    to alleviate technical obstacles, such as limited hardware keys and the
    need to protect a wide variety of BPF helper functions. We implement \tool 
    \ \mr{on Linux (ver. 6.1.38)}, and our evaluation shows that \tool 
    delivers 
    low-cost
    isolation of BPF programs under mainstream use cases, such as isolating a
    BPF packet filter with only 3\% throughput loss.
		
	\end{abstract}
	\acrodef{PMP}{Physical Memory Protection}
	\acrodef{BPF}{Berkeley Packet Filter}
	\acrodef{VM}{Virtual Machine}
	\acrodef{JIT}{Just-In-Time}
	\acrodef{DoS}{Denial of Service}
	\acrodef{OOB}{Out of Bound}
	\acrodef{MPK}{Memory Protection Keys}
	\acrodef{SVF}{Static Value-Flow}
	\acrodef{PTE}{Page Table Entry}
	\acrodefplural{PTE}{Page Table Entries}
	\acrodef{PKR}{Protection Key Register}
	\acrodef{MSR}{Model Specific Register}
	\acrodef{SAFE}{Scalability, Availability, Flexibility, Extensibility}
	\acrodef{LRU}{Least Recently Used}
	\acrodef{LKM}{Loadable Kernel Module}
	\acrodef{CFG}{Control Flow Graph}
	\acrodef{MTE}{Memory Tagging Extension}
	\acrodef{PMP}{Physical Memory Protection}
	\acrodef{IDT}{Interrupt Descriptor Table}
	\acrodef{GDT}{Global Descriptor Table}
	\acrodef{PKS}{Protection Key Supervisor}
	\acrodef{PKU}{Protection Key User}
	\acrodef{DACR}{Domain Access Control Register}
	\acrodef{SFI}{Software Fault Isolation}
	\acrodef{RCU}{Read-Copy Update}
	\acrodef{ROP}{Return-oriented Programming}
	\acrodef{PCID}{Process Context Identifier}
	\acrodef{TLB}{Translation Lookaside Buffer}

	\section{Introduction} \label{sec:intro}

It is common to extend kernel functionality by allowing user applications to
download code into the kernel space. In 1993, the well-known \ac{BPF} was
introduced for this purpose~\cite{bpfdoc}. The classic \ac{BPF} is an
infrastructure that inspects network packets and decides whether to forward or
discard them. With the introduction of its extended version (referred to as
eBPF) in the Linux kernel, \ac{BPF} soon became more powerful and is now
utilized in numerous real-life scenarios, such as load balancing, system tracing,
and system call filtering~\cite{bcc,xrp,bpfanaly,cilium,bpfnet1,bpfnet2}.

To ensure security, \ac{BPF} is equipped with a
\textit{verifier}~\cite{bpfverifier}. The verifier performs a variety of static
analyses to ensure the user-supplied code is secure. 
For instance, the verifier tracks the bounds of all pointers to prevent
out-of-bound access. Given that \ac{BPF} code runs directly within the kernel,
the verifier becomes crucial for \ac{BPF} security. Nevertheless, as pointed
out by recent studies~\cite{formaljit,jitk,prevail,prsafe,exobpf}, the
current verifier has various limitations and is insufficient for
the overall security of \ac{BPF}. \mr{First, the current \ac{BPF} ecosystem
  supports a variety of functionalities, such as packet forwarding and kernel
  debugging~\cite{katran,kprobe}. Supporting all these functionalities in the
  verifier results in a complicated verification process. Though the
  verifier has been partially verified via formal methods~\cite{formalbpf}, the
unverified part and the gap between abstraction and implementation still
result in 
vulnerabilities~\cite{CVE-2020-27194,CVE-2020-8835,CVE-2021-31440,CVE-2021-33200,CVE-2021-3444,CVE-2021-3490,CVE-2021-45402,CVE-2022-2785}}.
Second, due to the rapid expansion of BPF capabilities, the verifier is
frequently updated, and it is inherently difficult to update a complex static
verification tool without introducing new
vulnerabilities~\cite{CVE-2022-23222}.
To date, the \ac{BPF} subsystem has been repeatedly exploited. For instance,
two privilege-escalation vulnerabilities have been discovered in
\verb*|bpf_ringbuf|, a recent \ac{BPF} feature introduced in
2020~\cite{bpfdoc}. Further, the verifier's register-value tracking is quite
complex and often bypassed via corner-case operations~(e.g., sign
extension)~\cite{CVE-2020-27194,CVE-2020-8835,CVE-2021-31440,CVE-2021-33200}. 

\begin{figure*}[!htbp] \centering
  \includegraphics[width=0.80\linewidth]{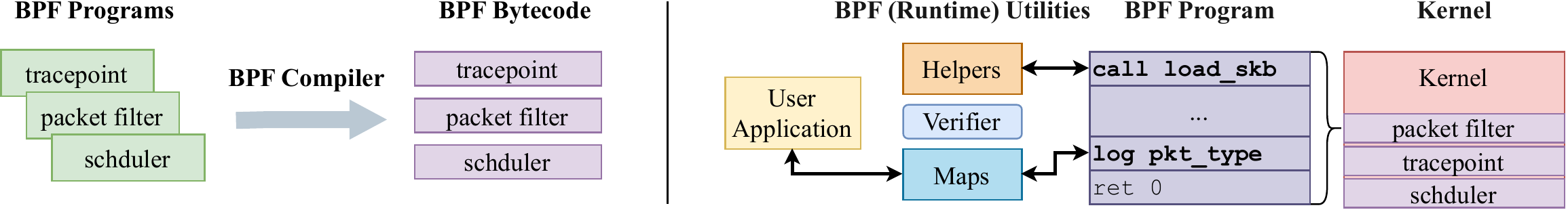} 
  \vspace{-3pt}
  \caption{BPF overview. We illustrate the BPF compilation procedure and
    execution context of a sample BPF packet filter.
}
\label{fig:bpf_overview} \vspace{-10pt} \end{figure*}

Given the increasing security threats in \ac{BPF} and the challenge of
enforcing safe \ac{BPF} programs with merely static verification, we seek to
employ hardware extensions to sandbox untrusted BPF programs. In particular, we
leverage Intel \ac{MPK}~\cite{sdm}, an emerging hardware extension that
partitions memory into distinct permission groups by assigning up to 16 keys to
their \acp{PTE}. With the aid of \ac{MPK}, we present \tool, which isolates
untrusted \ac{BPF} programs in a low-cost and principled manner. For instance,
two \ac{MPK} protection keys $K$ and $E$ can be assigned to the kernel and
the \ac{BPF} programs, respectively. When the kernel transfers control to a
\ac{BPF} program, it can set $K$ as access-disabled to prevent the potentially
malicious \ac{BPF} program from tampering with kernel memory.

Despite its promising potential, using \ac{MPK} to enforce \ac{BPF} isolation
is not straightforward. In designing \tool, we faced and overcame \textit{two
major technical hurdles}. 
First, \ac{MPK} provides a maximum of 16 keys. Thus, supporting numerous
\ac{BPF} programs with this limited number of keys is challenging. Existing
workarounds like key virtualization~\cite{libmpk} heavily rely on scheduling
and notification mechanisms that are only available to user space; our
tentative observation shows that directly reusing them in the kernel may
largely block kernel threads. To address this hurdle, we propose a novel
two-layer isolation scheme to protect both the kernel and the benign \ac{BPF}
programs from malicious \ac{BPF} programs. 
\tool also utilizes a contemporary hardware feature named process context
identifier to minimize the incurred overhead.
Second, while \ac{MPK}-based isolation mitigates malicious \ac{BPF} programs,
\textit{helper functions} provided by the \ac{BPF} subsystem may still be
exploited by attackers.
On the one hand, \tool\ should allow benign \ac{BPF} programs to use these
helpers freely. On the other hand, \tool\ must be cautious enough with these
helpers to ensure that they are not abused. \mr{However, designing security
policies for each of them requires non-trivial engineering effort and might
result in a bloated codebase. To prevent abuse, we design two defense schemes 
that do not rely on the specific design of helper functions. We show that each 
of
them applies to a wide range of helpers~(see
Appendix.~\hyperref[app:helper]{C}).}

We systematically examine how \tool\ mitigates the attack on the BPF ecosystem and the potential threats to \tool itself. We also empirically analyze all recent CVEs within \tool's
application scope. The result shows that \tool successfully mitigates each CVE.
\fixme{We evaluate the performance overhead brought by \tool across a variety of
micro and macro benchmark settings, and \tool\ achieves low performance
overhead across all settings. In particular, we evaluate \tool with the common use cases of \ac{BPF} in network
applications, and the maximum performance penalty from \tool is 3\% among
these cases. We also test \tool's overhead on system tracing,
another important \ac{BPF} use case. On average, \tool brings a performance
loss of 5.5\% in this setting. Furthermore, we evaluate \tool's performance when
using \ac{BPF} for system call filtering~\cite{seccomp}. In this case, the
performance loss brought by \tool is less than 3\%. Thus, we conclude that
\tool's overhead is reasonably low, especially given its security benefits.
In sum, we have made the following contributions.}
\begin{itemize}[leftmargin=*,labelindent=0pt,itemindent=0pt,parsep=0pt] \item Instead of
  merely relying on the BPF verifier to statically validate BPF 
  programs, this paper, for the first time, advocates isolating
  BPF programs with an emerging hardware extension, Intel MPK, \mr{effectively 
  ensuring the memory safety of BPF programs.}

  \item \mr{Technically, \tool is properly designed to address domain-specific
    challenges, including limited hardware keys and preventing helper abuse in
    the \ac{BPF} ecosystem. \tool\ features a two-layer isolation scheme to
    protect both the kernel and the benign \ac{BPF} programs from malicious
    \ac{BPF} programs and incorporates various design considerations to deliver
    a security guarantee on memory safety at a low cost.}

  \item We implemented a prototype of \tool on Linux 6.1.38\footnote{We release
    the codebase of \tool on our site \cite{moatrepo}. We will maintain \tool\
    to benefit the community and follow-up research.} and thoroughly evaluated
    its security over different attack scenarios~(including all memory-relevant 
    BPF CVEs in the past decade) and performance using various 
    benchmark datasets.
    The evaluation shows that \tool\ delivers a principled security warranty
    with minimum overhead.
\end{itemize}

	\section{Background}
\label{sec:bg}

\subsection{Berkeley Packet Filter (BPF)}
\label{subsec:bpf-background}

\parh{BPF Overview.}~\ac{BPF}~\cite{bpfdoc} was originally introduced to
facilitate flexible network packet filtering. Instead of inspecting packets 
in the user space, users can provide BPF instructions specifying packet filter
rules, which are directly executed in the kernel. BPF allows configurable
packet filtering without costly context switching and data copying.
Modern Linux kernel features extended BPF (eBPF), a Linux subsystem which
supports a wide range of use cases, such as kernel profiling, load balancing,
and firewalls.
\fixme{Popular applications such as Docker~\cite{docker}, Katran~\cite{katran},
and kernel debugging utilities like Kprobes~\cite{kprobe} utilize or are
built directly on top of BPF.}

\F~\ref{fig:bpf_overview} depicts an overview of how BPF programs are compiled
and deployed. The BPF subsystem offers ten general-purpose 64-bit
registers, a stack, BPF customized data structures (often called
BPF maps), and a set of BPF helper functions. To use BPF (e.g., for system tracing), 
users first write their own \ac{BPF} programs (in C code) to
specify the functionality, which, in turn, are compiled into
bytecode and loaded into the kernel. Given that BPF code is written by
untrusted users, the kernel employs a verifier to conduct several checks during
the bytecode loading stage (see below). By default, the verified bytecode is further
compiled into native code by an in-kernel \ac{JIT} compiler for better performance.
Additionally, on platforms without the \ac{JIT} support, the bytecode is alternatively
executed by the BPF interpreter.
The \ac{BPF} program is then attached to certain kernel
components based on its specific end goal. \fixme{For instance, as shown in
\F~\ref{fig:bpf_overview}, a \ac{BPF} program attaches to the kernel as the packet
filter, monitoring network traffic and sending statistics back to the user space via a BPF map.}

\begin{figure}[!htbp] 
	\centering
	\vspace{-4pt}
	\includegraphics[width=.9\linewidth]{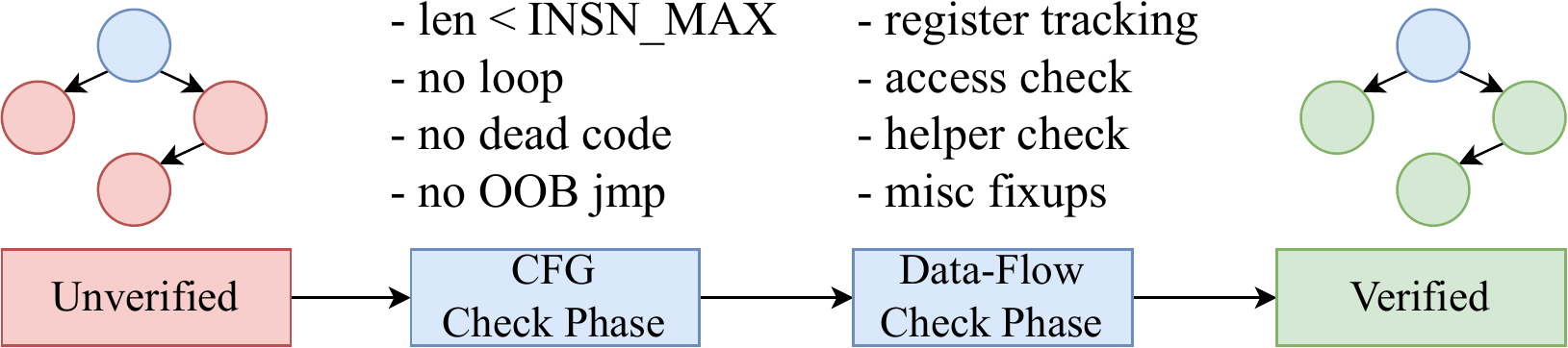}
	\caption{BPF verification process.}
	\label{fig:bpf_verifier}
	\vspace{-6pt}
\end{figure}

\parh{BPF Verifier.}~BPF programs are written in C and compiled into a
RISC-like instruction set. As aforementioned, the kernel strictly verifies the
BPF programs upon loading to ensure they are safe to execute.
\F~\ref{fig:bpf_verifier} illustrates the verification process in a holistic
manner. First, a \ac{BPF} program is parsed into a control flow graph (CFG) by
the verifier, which performs a CFG check phase to ensure four key
properties: 1) the program size is within a limit; 2) there are no back edges
(loops) on its CFG; 3) there is no unreachable code; and 4) all jumps are
direct jumps and refer to a valid destination. 

The verifier then tracks
the value flow of every register to deduce its value ranges conservatively.
With these ranges, the verifier decides if a pointer accesses safe memory and
if a parameter is valid. Since this analysis is performed statically, it is possible for a malicious \ac{BPF} program to exploit a vulnerability
to bypass it~\cite{CVE-2020-27194,CVE-2020-8835,CVE-2021-31440,CVE-2021-33200,CVE-2021-3444,CVE-2021-3490,CVE-2021-45402,CVE-2022-2785}.

\parh{BPF Helpers.}~The kernel also limits the functions a \ac{BPF} program
may call. Those functions are dubbed \ac{BPF} helpers, as shown in
\F~\ref{fig:bpf_overview}. To date, there are over 200 helpers provided by the kernel~\cite{helperlist}. Depending on the task, a \ac{BPF} program can usually call a group of relevant helpers.
For example, a \ac{BPF} packet filter can call \texttt{skb\_load}
to read packet data, but is not allowed to call any helper related to system tracing.

\parh{BPF Maps.}~Out of security concern, the kernel also sets a strict
space limit on \ac{BPF} programs. Each program, by default, can only use up to
512 bytes of stack space and 10 registers, which is far from enough for certain
\ac{BPF} programs. To address this problem, \ac{BPF} maps can be allocated to
provide additional space for \ac{BPF} programs. To date, there are over 30
types of maps supported by kernel~\cite{bpfmap}. Based on the isolation
requirements, they can be roughly categorized into two types. The first type is
maps that own a memory region. The most commonly used maps, hash maps and array maps,
belong to this category. \ac{BPF} programs use them to store data and
communicate with user space. Therefore, a proper access permission has to be
set for these maps \fixme{(see \tool's solution in \S~\ref{subsec:bpf-runtime}).} The second type holds references to other kernel resources~(e.g., file descriptors).
\ac{BPF} programs are restricted to using helpers to interact with this
type of map. Thus, \tool forbids \ac{BPF} programs from directly accessing them.

\begin{figure}[!htbp] 
	\centering
	\vspace{-5pt}
	\includegraphics[width=.85\linewidth]{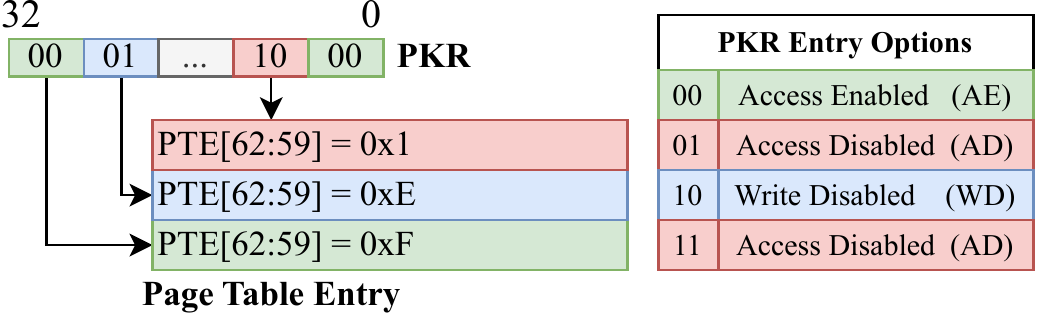}
	\vspace{-5pt}
	\caption{Intel MPK overview.}
	\label{fig:mpk}
	\vspace{-5pt}
\end{figure}

\parh{Classic \ac{BPF}~(cBPF).} \fixme{cBPF specializes in tasks like
  syscall filtering (e.g., \texttt{seccomp-BPF}) and has more restrictions than eBPF. We clarify that \tool supports both of them. We
  also evaluate \tool using \texttt{seccomp-BPF} with cBPF in \S~\ref{subsec:macro_bench}. In this paper, we use
\ac{BPF} to refer to both cBPF and eBPF, as the kernel internally converts cBPF
to eBPF.} 

\subsection{Hardware Features in \tool} \label{subsec:bg-hardware} 

\parh{Intel \ac{MPK}.}~Intel introduced \ac{MPK}~\cite{sdm} to provide
efficient page table permissions control. By assigning an MPK protection key to
the page table entries (PTEs) of one process, users can enable intra-process
isolation and confidential data access
control~\cite{erim,sgxlock,enclavedom,libmpk}. As illustrated in
\F~\ref{fig:mpk}, \ac{MPK} uses four reserved bits \texttt{[62:59]} in each
\ac{PTE} to indicate which protection key is attached to this page. Those three
\acp{PTE} in \F~\ref{fig:mpk} are assigned with keys \texttt{0x1}, \texttt{0xE}
and \texttt{0xF}, respectively. Since there are only 4 bits involved, the
maximum number of keys is 16. Then, a new 32-bit register named \ac{PKR} is
introduced to specify the access permission of these protection keys. Each key
occupies two bits in \ac{PKR}, whose values flag the access permission of the
page. In \F~\ref{fig:mpk}, the access permissions of the three pages are
\texttt{01} access-disabled~(AD), \texttt{10} write-disabled~(WD), and
\texttt{00} access-enabled~(AE), respectively.
By writing to certain bits in \ac{PKR}, the access permission of corresponding
pages can be configured efficiently without having to modify the PTEs.

\parh{Clarification and Notations.}~There are actually two
versions of \ac{MPK}. One applies to the user space, while the other
applies to the kernel space. For brevity, we refer to these two versions in
their conventional abbreviations as \ac{PKS} and \ac{PKU}, respectively. Most
existing works~\cite{libmpk,iskios,erim,sgxlock,enclavedom} are based on
\ac{PKU}. In \tool, we use \ac{PKS} instead since our goal is to isolate
in-kernel \ac{BPF} programs. The \fixme{logistics}
behind these two versions are mostly identical with slight variations. For
instance, the permission configuration register in \ac{PKS} is a \ac{MSR} named
\texttt{IA32\_PKRS}, which is inaccessible from user space, whereas in \ac{PKU},
this role is assigned to a dedicated register \texttt{PKRU}. To avoid confusion, the
rest of the paper refers to \ac{MPK} leveraged by \tool\ as \ac{PKS}.

\begin{figure}[htbp] 
	\centering
	\vspace{-5pt}
	\includegraphics[width=.8\linewidth]{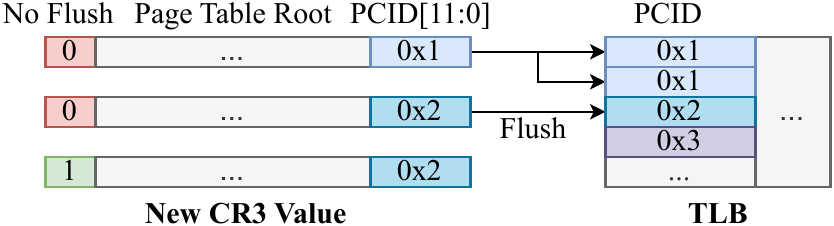}
	\caption{PCID overview.}
	\label{fig:pcid}
	\vspace{-5pt}
\end{figure}
\parh{\ac{PCID}.}~\fixme{\tool\ uses PCID to reduce the overhead of
\mbox{address-space} switching (\S~\ref{subsec:layer2}); we introduce PCID
here.}
On the x86 platform, the CR3 register holds the page table root of the current
process. Modifying the CR3 register causes a complete \ac{TLB} flush and is therefore costly. Fortunately, Intel introduced \ac{PCID} to address this
issue. As shown in \F~\ref{fig:pcid}, the lower 12 bits \texttt{[11:0]} of CR3
register are PCID, identifying the owner of the page table, while the highest
bit of the new CR3 value controls the flushing behavior of \ac{TLB}. If the
highest bit is 1, this modification does not flush \ac{TLB} at all; if the
highest bit is 0, then this modification only flushes the \ac{TLB} entries of
the \ac{PCID} in this new CR3 value. This feature enables fast address-space
switch without costly \ac{TLB} flush. Since there are only 12 reserved bits for
\ac{PCID}, it supports up to 4096 different processes with isolated \ac{TLB}
entries.

	\section{Motivation and Threat Model}
\label{sec:motivation_tm}

\subsection{Motivation}
\label{subsec:motivation} 
\mr{In this section, we discuss the typical threats to the BPF verifier, the
restriction on unprivileged BPF brought by these threats, and lastly, the
motivation for our research.}

\parh{Fast Feature Evolving.}~As a fast-developing technology, threats may come
from the inconsistency between the constantly expanding \ac{BPF} capabilities
and the rigorous static verification process imposed on
them~\cite{CVE-2022-23222,CVE-2021-3444}. It is a common practice to add
corresponding verification procedures simultaneously when introducing new
features to \ac{BPF} programs. However, it is difficult to implement a verifier
that supports all these features yet still does not miss any edge case, which
already has over 10K LoC with various functionalities~\cite{bpfverifier}.

\parh{Challenging Register-value Tracking.}~Second type of threats originates
from the complexity of the register-value tracking. \mr{Although the soundness of
such a tracking mechanism is formally proved~\cite{formalbpf}, there exist
gaps between the actual implementation and abstraction of the register-value 
tracking,
especially in some corner cases, such as sign extension, truncation, and bit
operators~\cite{CVE-2020-27194,CVE-2020-8835,CVE-2021-31440,CVE-2021-33200,CVE-2021-3444,CVE-2021-3490,CVE-2021-45402,CVE-2022-2785}.}

\parh{Unprivileged BPF.}~\mr{\ac{BPF} was originally designed as a restricted
  interface for \textit{unprivileged} users to extend kernel functionality. It
  comes with a fine-grained privilege system~\cite{bpfcap} that allows users to
  tune a specific part of the kernel without root. However, numerous
  vulnerabilities~\cite{CVE-2020-27194,CVE-2020-8835,CVE-2021-31440,CVE-2021-33200}
  indicate that the verifier is not reliable, and consequently, major
  distributions have banned unprivileged users from loading \ac{BPF}
  programs~\cite{suse,ubuntu}. Despite this, there is still a long-lasting 
  desire for unprivileged BPF in the community. For example, the 
  \texttt{seccomp-BPF}
  users have been asking for unprivileged BPF support for a long
  time~\cite{lwnseccomp}.\footnote{We clarify that \texttt{seccomp} already 
  supports classic-BPF
  	(cBPF), which lacks expressiveness and no longer updates~\cite{lwncbpf}. BPF
  	here refers to eBPF.} Moreover, there have been continuous
efforts~(from 2016 to 2023) in the community to re-emerge unprivileged BPF
again~\cite{lwneffort1,lwneffort2,lwneffort3}. Unfortunately, these efforts
fail as they only focus on enhancing the verifier itself, which is already
over-complicated and error-prone. 

\parh{Motivation.} Overall, seeing BPF's potential and its current restriction,
we propose \tool as an isolation scheme complementary to the BPF verifier. On
the one hand, this isolation scheme shall make \ac{BPF} more accessible to
unprivileged users whilst maintaining security. On the other hand, even for
privileged users, \tool provides the security guarantee that the BPF programs
obtained from a potentially untrusted source are isolated from the kernel.
Overall, we aim to provide a more secure and accessible BPF ecosystem, 
thereby promoting its development and adoption in the community.}

\subsection{Threat Model}
\label{subsec:threat-model}

Our threat model considers a practical setting that is
aligned with existing \ac{BPF}
vulnerabilities~\cite{CVE-2020-27194,CVE-2020-8835,CVE-2021-31440,CVE-2021-33200,CVE-2021-3444,CVE-2021-3490,CVE-2021-45402,CVE-2022-2785}.
Attackers can load their prepared \ac{BPF} code into the
kernel space to launch exploitation.
In particular, we assume attackers are \textit{non-privileged users} with
\ac{BPF} access since a root user already has control over almost the
entire kernel. 
\tool\ isolates user-submitted BPF programs
and prevents them from accessing kernel memory regions. 
\fixme{As will be introduced in \S~\ref{sec:design}, a BPF program is given
only the necessary resources and privileges to complete its task. We present
the threat models of major components in our research context as follows.}

\parh{\ac{BPF} Programs.}~We assume that
malicious \ac{BPF} programs are able to bypass checks statically performed by
the verifier; they may thus behave maliciously during runtime. Our threat model
deems BPF programs as \textit{untrusted}. 

\parh{\ac{BPF} Helper Functions.}~These helpers act as the intermediate layer
between the \ac{BPF} subsystem and kernel. Certain malicious BPF
programs can abuse these helpers to perform attacks, and therefore, we assume
they are also \textit{untrusted}. \tool mitigates risks raised by
adversarial-manipulated helper functions with practical defenses.

\parh{Out of Scope.}~\fixme{The main objective of \tool is to mitigate memory
  exploitation performed by \ac{BPF} programs. Other subtle attacks (not
  relevant to memory exploits), such as speculation, race condition, and
  \ac{DoS} toward the BPF subsystem~\cite{cve_race,cve_spec} are not
  considered. They do not specifically exist in BPF~\cite{sgxpectre,smashex},
  and are addressed by relevant research~\cite{dosdetect,specdef}. We thus
  treat them as orthogonal. Also, \ac{BPF} subsystem comes with a set of
  user-space facilities such as \texttt{libbpf}; bugs in them are not
  considered by \tool. \mr{Note that \tool mitigates information leakage that
  is due to out-of-bounds memory access; if the leakage is due to issues
  like speculation~\cite{cve_spec}, then it is out of the 
  scope of \tool.}

We clarify that \tool focuses on the kernel memory exploitation via \ac{BPF},
its most prevalent threat. The vulnerabilities mitigated by \tool typically
receive high threat scores in vulnerability
databases~\cite{CVE-2020-27194,CVE-2020-8835,CVE-2021-31440,CVE-2021-33200,CVE-2021-3444,CVE-2021-3490,CVE-2021-45402,CVE-2022-2785,CVE-2022-23222}
with public PoC exploits~\cite{exp23222}, whereas above-precluded
vulnerabilities often lack exploits~\cite{CVE-2020-27171}.}

	\section{Design}
\label{sec:design}

\parh{\tool Overview.}~As described in \S~\ref{subsec:motivation}, the current
security design against malicious \ac{BPF} programs solely relies on the static
analysis performed by the \ac{BPF} verifier, which is seen as a weak point and
exploitable by non-privileged users. \tool\ instead delivers a principled
isolation of \ac{BPF} programs from the rest part of the kernel using \ac{PKS} 
and prevents bypasses. 

\begin{figure}[!htbp]
	\vspace{-4pt}
	\centering
	\includegraphics[width=0.85\linewidth]{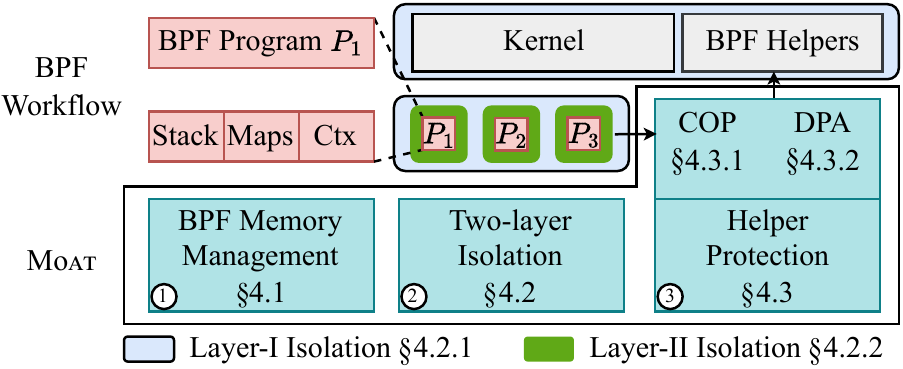}
	\vspace{-4pt}
	\caption{\tool overview.}
	\label{fig:overview}
	\vspace{-5pt}
\end{figure}

\F~\ref{fig:overview} depicts an overview of \tool and how it is
integrated into the workflow of \ac{BPF} programs. \circled{1} Given a
user-submitted BPF program $P$, \tool\ statically allocates the necessary
memory regions the program needs, such as stack, maps, and context based on
$P$'s metadata~(\S~\ref{subsec:bpf-runtime}). \circled{2} When the kernel
invokes $P$, \tool isolates $P$ from the kernel using \ac{PKS}~(Layer-I in
\S~\ref{subsec:layer1}), and constrains $P$ in its isolated address
space~(Layer-II in \S~\ref{subsec:layer2}). \circled{3} \fixme{On the occasions
that $P$ calls helpers, depending on the helper types, \tool adjusts the
involved memory region permissions~(\S~\ref{subsec:cop}) and validates
the helper parameters~(\S~\ref{subsec:dpa}) to prevent the helpers from being
abused.}

\begin{table*}[htbp]
	\caption{\ac{BPF} context of common program types.}
	\vspace{-5pt}
	\label{table:bpf_ctx}
	\centering
	\resizebox{0.8\linewidth}{!}{
\begin{tabular}{lllccl}
	\hline
	\textbf{Category}                 & \textbf{Program Type} & \textbf{Context Type}         & \textbf{Persistent} & \textbf{Nested} & \textbf{Note}                                \\ \hline
	\multirow{4}{*}{\textbf{Network}} & Socket Filter         & \texttt{sk\_buff *}                    & Yes                 & Yes             & Socket packet buffer                         \\ \cline{2-6} 
	& Socket Ops            & \texttt{bpf\_sock\_ops *}              & No                  & Yes             & Socket events (timeout, retransmission, ...) \\ \cline{2-6} 
	& Socket Lookup         & \texttt{bpf\_sk\_lookup *}             & No                  & Yes             & Packet information for socket lookup         \\ \cline{2-6} 
	& XDP                   & \texttt{xdp\_md *}                     & No                  & Yes              & Metadata of \texttt{xdp\_buff}                        \\ \hline
	\multirow{3}{*}{\textbf{Tracing}} & Kprobe                & \texttt{pt\_regs *}                    & No                  & No              & Register status on probed location           \\ \cline{2-6} 
	& Tracepoints           & Depending on tracepoint types & No                  & No              & Relevant tracepoint information              \\ \cline{2-6} 
	& Perf Event            & \texttt{bpf\_perf\_event\_data *}      & No                  & No              & Perf. event (register status, sample period) \\ \hline
	\multirow{2}{*}{\textbf{Cgroup}}  & Cgroup Socket Filter  & \texttt{sk\_buff *}                    & Yes                 & Yes             & Socket packet buffer under specific cgroup   \\ \cline{2-6} 
	& Cgroup Device         & \texttt{bpf\_cgroup\_dev\_ctx *}      & No                  & No              & Device ID, access type (read, write)         \\ \hline
\end{tabular}
	}
	\vspace{-12pt}
\end{table*}
\subsection{\ac{BPF} Memory Management in \tool}
\label{subsec:bpf-runtime}
Further to the overview in \F~\ref{fig:overview}, we introduce how \tool 
manages the \ac{BPF} memory.
The \ac{BPF} memory refers to the memory regions a \ac{BPF} program needs to function properly, including descriptor tables, stacks, maps, and runtime context.

\parh{Descriptor Tables.}~On x86 platforms, \ac{GDT} and
\ac{IDT} are essential for basic operations like interrupt. These structures
are assigned to a shared region that all \ac{BPF} programs can access. To
prevent tampering, they are made read-only when shared.

\parh{Stack.}~\ac{BPF} programs use a 512-byte stack space to store local
variables and function frames. The verifier determines if a program makes
out-of-bounds access toward the stack. Thus, if the \ac{BPF} program passes the
static checks, its stack is directly allocated from the kernel stack. However,
as discussed in \S~\ref{subsec:motivation}, certain vulnerabilities may allow
\ac{BPF} programs to bypass this check. Thus, \tool needs to allocate the stack
as a part of \ac{BPF} memory and swap stacks to prevent the \ac{BPF} programs from
tampering with the kernel stack.

\parh{Maps.}~As described in \S~\ref{subsec:bpf-background}, maps are utilized by \ac{BPF}
programs to store data and communicate with the user space. Linux provides a set of
helper functions for \ac{BPF} programs to interact with maps. For example,
\texttt{bpf\_map\_look\_up\_elem} returns the pointer of an element so that
the program can modify its value. This means that \ac{BPF} programs must have
access to these elements' memory. Thus, \tool allocates these maps as a part of
\ac{BPF} memory. Note that we do not allocate the metadata of these maps inside
\ac{BPF} memory since they contain exploitable structures like function
pointers.

\parh{Runtime Context.}~The context refers to \ac{BPF} program parameters,
which vary depending on the \ac{BPF} program types. We investigated the \ac{BPF}
contexts of common \ac{BPF} program types and summarized our findings in
\T~\ref{table:bpf_ctx}. Most of these contexts are local objects on the
kernel stack and are passed to \ac{BPF} programs as parameters, such as
\texttt{bpf\_cgroup\_dev\_ctx}. For this type of \ac{BPF} context, \tool
allocates them on the \ac{BPF} stack instead so that the \ac{BPF}
programs can still access them without the permission to access the kernel
stack. However, there also exist contexts that are not local objects on the stack
but persistent kernel structures. For example, \texttt{sk\_buff} holds the
network packet received by a socket and is also passed to \ac{BPF} socket
filter programs as context. For this type of persistent context~(denoted in the
fourth column of \T~\ref{table:bpf_ctx}), \tool dynamically maps the physical
page of the corresponding context into the \ac{BPF} memory. The reason why we
choose to map instead of creating a local copy is that \texttt{sk\_buff}
is typically hundreds of bytes. Our preliminary experiment shows
that syncing between the local copy and the actual kernel object brings
non-trivial overheads. Furthermore, network-related \ac{BPF} contexts~(e.g.,
\texttt{bpf\_sock\_ops}) may contain nested pointers to other kernel
structures~(denoted in the fifth column of \T~\ref{table:bpf_ctx}).
Including only these pointers in \ac{BPF} memory triggers a false alarm, as these
nested structures are not included. We clarify that \ac{BPF} programs only
access limited fields of these nested structures. Thus, \tool reserves a part
of \ac{BPF} memory to mirror these nested fields efficiently so that they can
be accessed by the \ac{BPF} programs.

\begin{figure}[!htbp]
	\centering
	\vspace{-5pt}
	\includegraphics[width=0.8\linewidth]{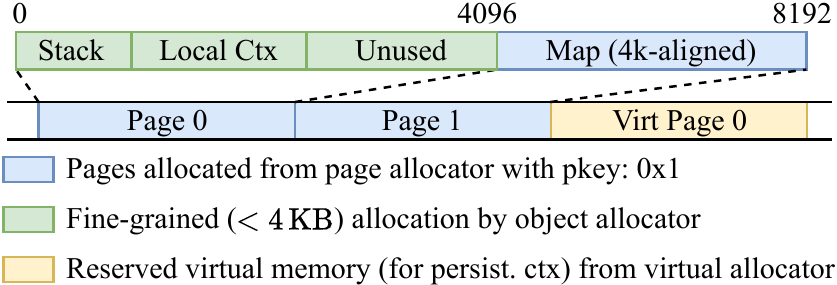}

	\caption{\ac{BPF} memory allocators.}
		\vspace{-5pt}
	\label{fig:bpf_mem}
\end{figure}

\parh{BPF Memory Allocators.}~As shown in \F~\ref{fig:bpf_mem}, \tool provides
three types of allocators to manage these \ac{BPF} memory regions. When loading
a BPF program, the page allocator first allocates physical pages~(Page 1) for
its \ac{BPF} memory; these pages are given the protection key \texttt{0x1} and
become a part of its \ac{BPF} memory. The object allocator handles fine-grained
allocations from the allocated pages~(Page 0) that are less than the page
size~(4\,KB), e.g., the BPF stacks~(512 bytes). Lastly, the virtual allocator
controls the virtual memory that is not backed with concrete physical pages. For
instance, it reserves a part of virtual \ac{BPF} memory~(Virt Page 0) to map
persistent \ac{BPF} contexts~(e.g., \texttt{sk\_buff}). Note that we also modify
the implementation of \ac{BPF} maps to use \tool's allocator.

\subsection{Two-layer Isolation}
\label{subsec:scalability}

\parh{Challenge.}~In theory, we can assign each \ac{BPF} program with a unique
key to achieve low-cost \ac{BPF} isolation. However, \ac{PKS} only supports up
to 16 regions~(i.e., keys). If we assign each \ac{BPF} program with a unique
key, these keys would soon be exhausted as there could be over 16 \ac{BPF}
programs in the kernel. It is challenging to isolate an unlimited number of
\ac{BPF} programs with only 16 protection keys.

\parh{Solution.}~We propose a novel two-layer
isolation scheme using \ac{PKS} and isolated address spaces. Though
isolating address spaces is less efficient than \ac{PKS}, we manage to reduce
its overhead to a minimum using a contemporary hardware feature named PCID; We use
PCID as a complement to \ac{PKS} to support the isolation of numerous \ac{BPF} programs.

\subsubsection{Layer-I: Lightweight Isolation Domain via \ac{PKS}}
\label{subsec:layer1} 

The main objective of \tool is to isolate the kernel from malicious \ac{BPF} programs. Thus, we use \ac{PKS} as a
lightweight isolation primitive between kernel and \ac{BPF}
programs. Specifically, we use \ac{PKS} to build three isolated domains: the \ac{BPF} domain, the kernel domain, and the shared domain. 

\begin{figure}[htbp]
	\centering
	\includegraphics[width=0.9\linewidth]{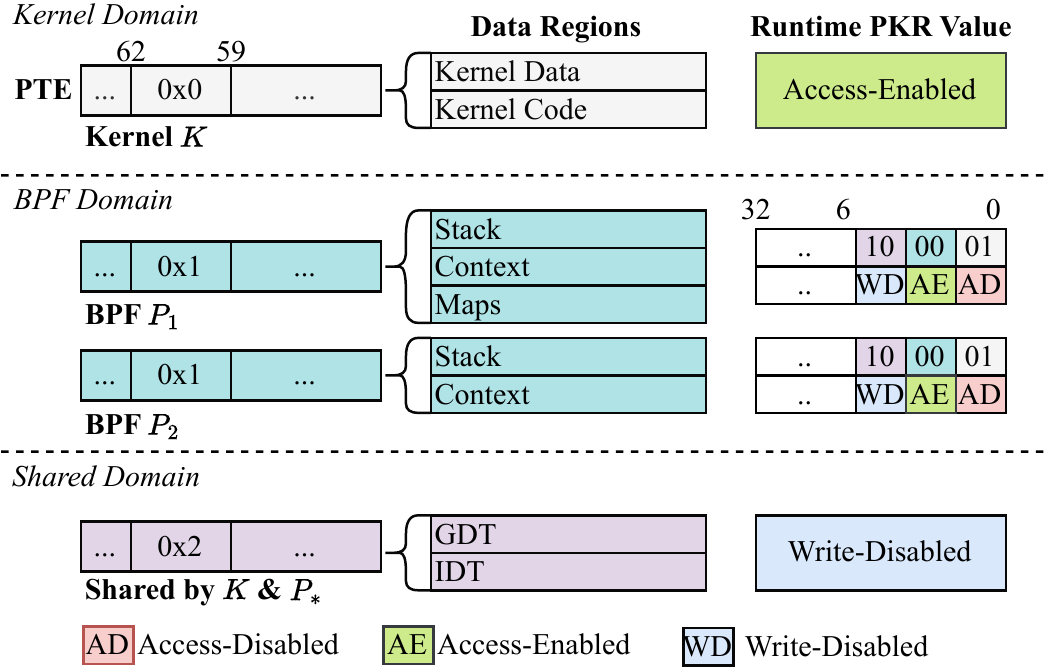}
	\vspace{-3pt}
	\caption{\ac{PKS}-enforced domains of \tool.}
	\vspace{-5pt}
	\label{fig:bpf_domains}
	\vspace{-5pt}
\end{figure}

\fixme{
As depicted in \F~\ref{fig:bpf_domains}, all \ac{BPF} programs
reside in the \ac{BPF} domain with the protection key \texttt{0x1}. \tool
grants a \ac{BPF} program access~(i.e., access-enabled; AE) to the BPF domain (\texttt{0x1}) when executing the program by setting its \ac{PKR} bits to
\texttt{00}~(flagging AE). The kernel domain holds all kernel
pages with the protection key \texttt{0x0} and is only
accessible by the kernel itself.  When entering a \ac{BPF} program, this
kernel domain (\texttt{0x0}) becomes access-disabled~(AD) by setting its \ac{PKR} bits to \texttt{01}~(flagging AD). However, the shared domain (\texttt{0x2})
comprises memory regions like \ac{IDT} and \ac{GDT}. These regions are crucial
for low-level routines like interrupts. Thus, they are made
write-disabled~(WD) instead of access-disabled for \ac{BPF} programs by setting the \ac{PKR} bits to \texttt{10}~(flagging WD).
}

This domain design only needs three (out of 16) keys from \ac{PKS} yet
effectively mitigates malicious \ac{BPF} programs targeting the kernel. 
For these malicious programs, a modus operandi is to introduce an unsanitized kernel pointer
by exploiting a verifier vulnerability. Then, the malicious \ac{BPF} program
arbitrarily tampers the kernel using that pointer, leading to a full-blown
exploitation. Isolating \ac{BPF} programs from the kernel
effectively stops such attacks, as all malicious kernel access directly from
\ac{BPF} programs is prevented by \ac{PKS}.

\subsubsection{Layer-II: Isolated \ac{BPF} Address Space} \label{subsec:layer2}

In \S~\ref{subsec:layer1}, we have discussed how \tool prevents \ac{BPF}
attacks targeting the kernel. However, \tool only uses \ac{PKS} to build
isolation between the kernel and \ac{BPF} programs. All \ac{BPF} programs still
share the same \fixme{PKS} domain, allowing a malicious \ac{BPF} program to tamper with the
memory of benign \ac{BPF} programs. 
Inspired by prior works in user-space isolation~\cite{vdom}, we set
up an isolated address space for each \ac{BPF} program to prevent such
tampering. Consequently, when a malicious \ac{BPF} program tries to access
the memory regions of another \ac{BPF} program, a page fault occurs, and the malicious
\ac{BPF} program is immediately terminated.

\begin{figure}[!htbp]
	\centering
	\vspace{-10pt}
	\includegraphics[width=0.9\linewidth]{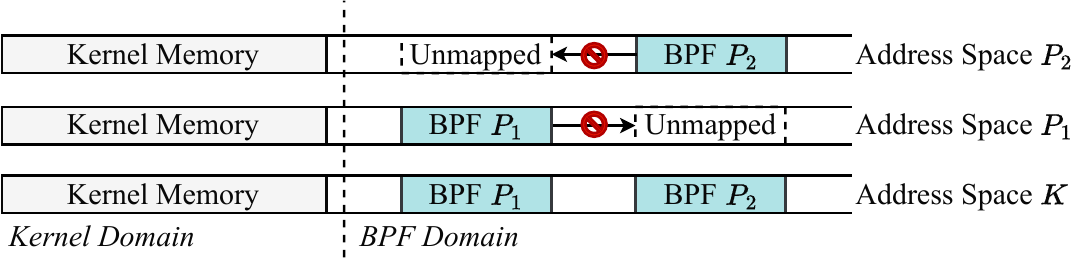}
	\vspace{-5pt}
	\caption{Isolated address spaces of \ac{BPF} programs.}
	\vspace{-5pt}
	\label{fig:isolated_addr}
\end{figure}

\F~\ref{fig:isolated_addr} illustrates the isolated address spaces of two
\ac{BPF} programs, $P_1$ and $P_2$. In the address space of $P_1$, the mapping
of $P_2$ does not exist. Similarly, the mapping of $P_1$ does not exist in the
address space of $P_2$ either. This effectively prevents \ac{BPF} programs from
accessing each other and addresses the abovementioned issue. 
To avoid the high \ac{TLB} flush overhead~(and TLB misses) from the
address-space switching, we use \ac{PCID} to keep the \ac{TLB} entries from
different \ac{BPF} programs isolated. Since \ac{BPF} programs are usually
smaller than user applications, \tool allocates a non-overlapped virtual
address space with a unique \ac{PCID} for each of them. In the rare cases where
over 4,096 \ac{BPF} programs are running in the same kernel, \tool has to flush
the TLB when two \ac{BPF} programs~(with the same PCID) run consecutively. Since there are 4,096 PCIDs available, we expect such conflicts to be rare, especially considering that the kernel
also periodically flushes \ac{TLB}, thus clearing the conflicting entries. Even
when such cases occur, \tool only flushes the TLB entries of the conflicting
PCID and leaves other TLB entries intact.

\parh{Why Intra-BPF Isolation.}~One may question the necessity for the
isolation between \ac{BPF} programs, as most existing \ac{BPF} exploits target
the kernel. However, since \ac{BPF} maps are the only bridge between the
\ac{BPF} programs and the user space, the configurations of a \ac{BPF} program
have to be saved in its maps so that users can change its behavior without
reloading it. This paradigm makes cross-\ac{BPF} attack a noticeable
threat~\cite{bpfconf}. For example, an attacker may load a malicious \ac{BPF}
program to change the behavior of another program by tampering with its
configuration maps, disrupting resources accounting, or even nullifying
security checks. Intra-BPF isolation is essential for preventing such attacks.

\subsection{Helper Security Mechanism}
\label{subsec:extensibility}

As mentioned in \S~\ref{subsec:bpf-background}, the kernel provides a set of
helper functions for \ac{BPF} programs. As these helpers act as the interfaces
between the kernel and \ac{BPF} programs, they can also be abused by malicious
programs to launch attacks. \tool needs to protect the helpers from such abuse.

Our investigation of existing BPF vulnerabilities shows that the malicious BPF
programs typically abuse the BPF helpers in two ways: \ding{192} the helper
contains a defect, which is exploited by the malicious 
programs~\cite{CVE-2021-34866}; \ding{193} the helper itself is
correct, but the malicious programs pass invalid parameters to abuse
it~\cite{CVE-2022-23222}. For \ding{192}, a typical case is that the helper
itself contains defects such as heap overflow. These defects are leveraged by
malicious programs to overwrite the sensitive fields~(e.g., function pointers)
of BPF-related kernel objects. For \ding{193}, since the helper by itself is
correct, the malicious programs are typically restricted to leaking kernel
pointers (by passing invalid parameters) and cannot conduct full-blown
exploitation. \mr{Notably, in most cases, since
  the helper parameters are checked by the \ac{BPF} verifier, the malicious
  programs still need to leverage the register-value-tracking vulnerabilities
  in the verifier (\S~\ref{subsec:motivation}) to bypass this
  check~\cite{CVE-2020-27194,CVE-2020-8835,CVE-2021-31440,CVE-2021-33200,CVE-2021-3444,CVE-2021-3490,CVE-2021-45402,CVE-2022-2785}.
}

\parh{Challenge.}~\mr{However, protecting these BPF helpers is not trivial. First,
the BPF helpers, by design, need to access BPF-related objects in the kernel
memory; blindly isolating helpers using PKS leads to spurious alarms and
impedes benign programs. Second, there are over 200 BPF helpers in the kernel,
so \tool's design must be generic enough to apply to most of these
helpers~(see Appendix.~\hyperref[app:helper]{C} for the supported helpers).}

\parh{Design Consideration and Solution.}~\mr{To prevent such abuse, we aim to
  identify and guard sensitive BPF-related objects (instead of all BPF-related
  objects) from the defective BPF helpers. Besides directly protecting
  sensitive objects, since the attackers need to deliver malformed parameters
  to conduct helper abuse, we also wish to ensure the validity of the helper
  parameters at runtime. To this end, we designed the following two defense 
  schemes: Critical Object Protection~(COP) and Dynamic
  Parameter Auditing~(DPA). COP protects sensitive BPF-related objects from
  being tampered with, while DPA dynamically
  checks if the arguments of the helpers are within legitimate ranges. To 
  clarify, COP and DPA should be enabled together to deliver protection. DPA 
  only constrains the arguments to their expected ranges. This stops most 
  exploitation 
  attempts~\cite{CVE-2020-27194,CVE-2020-8835,CVE-2021-31440,CVE-2021-33200,CVE-2021-3444}
   but may not ward them off completely in the presence of a buggy 
   helper~\cite{CVE-2021-34866}; COP, in this case, prevents the buggy helper 
   from 
   accessing 
   sensitive objects.}

\subsubsection{Critical Object Protection (COP)}
\label{subsec:cop}

\begin{figure}[htbp] \centering
		\vspace{-5pt}
	\includegraphics[width=0.75\linewidth]{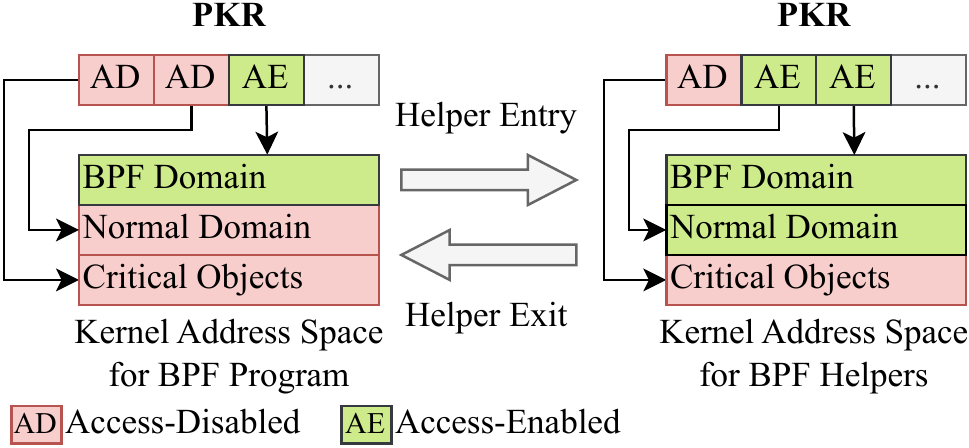} 
		\vspace{-5pt}
	\caption{Critical object protection (COP).} \label{fig:cop} 
		\vspace{-10pt}
\end{figure}

Although \ac{BPF} helpers have to access BPF-related kernel objects to complete
their tasks, the sensitive BPF-related objects should still not be accessed by
any helper. For example, \texttt{array\_map\_ops} is a function pointer in the
BPF array maps that should \textit{only} be accessed from system calls.
However, \texttt{array\_map\_ops} is close to other helper-needed objects in
the address space, making it a potential victim of the abused helpers.
Based on this, we designed the COP scheme. As shown in \F~\ref{fig:cop},
instead of treating the entire kernel domain as a whole, we divide it into a
normal domain and a critical-object domain. Permissions of these critical
objects are managed via \mr{an extra page and protection key}. When entering
helper functions, \fixme{instead of setting the entire kernel space as
access-enabled~(AE)}, those critical objects remain access-disabled~(AD),
preventing the helpers from accessing them.
\mr{To identify these objects, we first review BPF CVEs 
	and find all objects that have been exploited. Then, we manually 
	search for similar objects in the kernel and check that 
	these found objects indeed contain sensitive fields~(e.g., function 
	pointers). 
	It took 
	two authors about half a month to conduct the above procedure 
	individually and cross-check results, which ensures the credibility 
of our
research to a great extent.} \mr{We identified a total of 44 critical objects~(see
Appendix~\hyperref[app:cop]{A}); these objects could either leak the sensitive
base address of the kernel~(e.g., \texttt{iter\_seq\_info}) or even be tampered
with to launch a full-blown exploit~(e.g., \texttt{array\_map\_ops}). In
addition to these 44 identified objects, we set \tool itself and \texttt{cred}
as critical objects. The former contains sensitive data of \tool~(e.g., the
saved state of \texttt{IA32\_PKRS}), while the latter tracks the privilege of a
process.
We believe protecting the identified critical objects provides a \mr{practical}
security guarantee for the BPF helpers.} \mr{Nonetheless, unidentified 
critical objects could exist; we will discuss their potential threat in 
\S~\ref{sec:discussion}. Moreover, it is always feasible to
extend COP to protect other kernel objects.}

\subsubsection{Dynamic Parameter Auditing (DPA)} \label{subsec:dpa} 
\begin{figure}[htbp]
	\vspace{-3pt}
	\centering
	\vspace{-4pt}
	\includegraphics[width=0.95\linewidth]{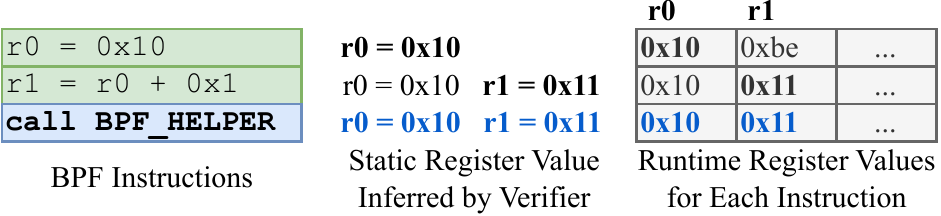}
	\vspace{-3pt}
	\caption{Register value tracking of the verifier.}
\label{fig:dynamic_auditing}
\vspace{-5pt}
\end{figure}

To further
regulate the helpers, we propose Dynamic Parameter Auditing (DPA), which
leverages the information obtained from the BPF verifier to dynamically check
if the parameters are within their legitimate ranges. As illustrated in
\F~\ref{fig:dynamic_auditing}, the verifier can deduce the value range of each
register via static analysis (\fixme{aligned with the uncovered verifier
  inaccuracies~\cite{CVE-2021-3490,CVE-2021-31440}; our DPA design tolerates
  even \textit{invalidly} deduced value ranges; see clarification below}).
  \tool logs such value ranges and instruments the BPF programs to insert checks before helper calls. During runtime, these checks ensure that the provided parameters of the helpers are within the verifier-deduced value ranges during runtime. In our
  example, we can check if \texttt{r0==0x10;r1==0x11} when \texttt{BPF\_HELPER}
  is called. If the parameter runtime values do not match with the static
  analysis results, the \ac{BPF} program is terminated immediately.

\noindent \underline{Clarification.}~In this DPA strategy,
  \fixme{one may wonder if the ``value ranges'' deduced by the
  verifier are wrong~\cite{CVE-2021-3490,CVE-2021-31440}.} To clarify this, we list possible cases of a BPF
  variable $v$'s value range and the corresponding system states in
\T~\ref{table:dpa_cases}; our discussions are as follows.

\begin{table}[htbp]
	\centering
	\caption{Four cases of a BPF variable $v$'s value ranges. $R$ denotes the
	runtime value of $v$, $D$ denotes the verifier's deduced value of $v$, $E$
	denotes verifier's \textit{expected} legitimate value range of $v$, while $T$
	denotes the \textit{ground truth} legitimate value range of $v$. The last column denotes this case is safe~(\ding{51}), mitigated by verifier~(\ding{51}\textsubscript{\textsc{v}}), mitigated by \tool~(\ding{51}\textsubscript{\textsc{m}}), or unsafe~(\ding{55})}
	\label{table:dpa_cases}
	\resizebox{0.75\linewidth}{!}{
	\begin{tabular}{c|ccccc}
		\hline
	       &  $R$ & $D$ & $E$ & $T$ & \textbf{State}       \\\hline
		1 & \texttt{0x10}             & \texttt{0x10}                    & \texttt{{[}0,0x20{]}}                                & \texttt{{[}0,0x20{]}}                & \ding{51} \\\hline
		2 & \texttt{\textbf{0xba}} & \texttt{0xba}        & \texttt{{[}0,0x20{]}}                    & \texttt{{[}0,0x20{]}}                & \ding{51}\rlap{\textsubscript{\textsc{v}}} \\\hline
		3 & \texttt{\textbf{0xba}} & \textbf{\texttt{0x10}}                    & \texttt{{[}0,0x20{]}}                                & \texttt{{[}0,0x20{]}}                & \ding{51}\rlap{\textsubscript{\textsc{m}}} \\\hline
		4 & \texttt{\textbf{0xba}} & \texttt{\textbf{0xba}}        & \texttt{\textbf{{[}0,0xba{]}}}                    & \texttt{{[}0,0x20{]}}                & \ding{55} \\\hline
	\end{tabular}}
	\vspace{-10pt}
\end{table}

Case 1 illustrates the value range of a variable $v$ in a benign \ac{BPF}
program. The runtime value aligns with verifier's deduction which further falls
within the \textit{expected} and \textit{true} legitimate value ranges
simultaneously~($R=D\in E=T$, see the caption of \T~\ref{table:dpa_cases}).
Case 2 demonstrates the value range of variable $v$ in a malformed \ac{BPF}
program. The runtime value \texttt{0xba} is out-of-bounds, and this invalid
value is detected by the verifier through static analysis. Therefore, this
program is rejected by the verifier, and the system remains safe~($R=D\notin
E=T$).
Case 3 shows the value range of $v$ in a malicious \ac{BPF} program. The
runtime value \texttt{0xba} is out-of-bounds. However, due to the incomplete
analysis caused by vulnerabilities, the verifier deduces that $v$'s value is
\texttt{0x10}, which is within the verifier's expectation. Since DPA operates
in the runtime and checks whether the runtime value actually matches the
verifier's deduction, this mismatch is then detected by DPA, and this malicious
\ac{BPF} program is terminated~($R\neq D\in E=T$).

While the above three cases cannot be exploited, Case 4 implies a scenario
where DPA fails and the helper is abused. The verifier's \textit{expected}
value range differs from the \textit{ground truth}, legitimate value range.
This discrepancy allows an out-of-bounds value \texttt{0xba} to be passed as an
argument to a helper for exploitation. For this to occur, the following
conditions must be satisfied simultaneously: \circled{1} The verifier has an
\textit{incorrect} expectation (i.e., $E\neq T$). \circled{2} The incorrect
expectation $E$ is \textit{unsafe} (i.e., $T-E$ overlaps an exploitable
structure). \circled{3} The BPF program is carefully tweaked to be aligned with
$D$ and evade DPA (i.e., $R=D$).
For \ac{BPF} programs, It is usually straightforward for the verifier to obtain
$E$ statically (e.g., $E$ encodes the array size). It is thus hard to satisfy
\circled{1} and \circled{2} simultaneously. For today's known \ac{BPF} exploits
(all of which fall into Case 3), the verifier has the correct expectation $E=T$
but makes the incomplete deduction $R\neq D$; therefore, the discrepancy $E\neq
T$ is never encountered in practice.

\subsection{Design Comparison} 
\label{subsec:design_comparison}

\mr{In this section, we compare \tool's design with other works in kernel
isolation. This helps highlight the contribution of \tool, in comparison to 
previous research.}

\parh{Virtualization.} \mr{There is a line of research
  works on isolating kernel components via virtualization~\cite{lxd,lvd,skee}.
  However, among these prior works, lightweight solutions like SKEE~\cite{skee}
  are not compatible with the low-level routines (e.g., interrupt) in Linux.
  SKEE disables the interrupt upon entry, but disabling the interrupt will 
  significantly
  impede BPF's network performance. To incorporate with low-level routines in
  the kernel, non-trivial modification to the system is often needed. For
  example, LVD and LXDs~\cite{lxd,lvd} require a hypervisor and an implanted
  micro-kernel to manage the isolated components.  \tool, on the one hand,
  leverages PKS to enforce lightweight isolation between kernel and BPF; this
ensures efficient interrupt handling. On the other hand, \tool re-uses
the kernel memory subsystem to enforce intra-BPF isolation without the 
additional hypervisor or micro-kernel.}

\parh{SFI-based Solutions.} \mr{Prior works also proposed Software Fault
  Injection~(SFI) for kernel security~\cite{sfiker,sandbpf}. SFI
  inserts checks before memory access to ensure that they fall into
  valid ranges. However, inserting checks for memory access
  often brings higher overheads~(see \S~\ref{subsec:add_eval} for an 
  empirical comparison with \tool). \tool uses PKS to ensure memory safety and 
  only inserts checks 
  before helpers. Since the number of helper calls is much smaller than 
  that of memory access, this design largely reduces the overhead of \tool.}

	\section{Implementation}
\label{sec:implementation}

\tool is implemented on Linux 6.1.38, and consists of \fixme{2,911} lines
of C code. We explain the key points below.

\parh{Kernel Interrupt Handling.}~\tool has to cooperate with many low-level
routines inside the kernel. For instance, during the execution of \ac{BPF}
programs, an interrupt may occur and take over the control flow to its handler.
Note that most interrupt handlers require access to kernel memory, and as a
result, the \ac{PKS} would presumably raise spurious alerts. Thus, we need to
temporarily disable \ac{PKS} inside these handlers and re-enable it
once the handlers finish. To avoid the overhead when there is no
\ac{BPF} program, we use a per-CPU variable \texttt{in\_bpf}
to identify whether the processor is executing a \ac{BPF} program. Since
\ac{BPF} programs only occupy a tiny fraction of kernel execution time, we
observe little performance loss due to this, even under cases
where interrupt frequently occurs~(e.g., intensive network activity in
\S~\ref{subsec:macro_bench}).

\parh{Granularity of \ac{PKS}.}~As protection keys are associated with PTEs,
\tool only protects memory in the granularity of a page~(i.e., 4\,KB). However,
the objects used by \ac{BPF} programs may not be aligned to 4\,KB, which means
they could interleave with critical kernel structures. Therefore, granting
\ac{BPF} programs access to these objects also enables access to those kernel
structures and leads to exploitation. To prevent this, we have modified
\ac{BPF}-related objects (e.g., maps) so that they are page-aligned and not
interleaved with other structures.

\begin{figure}[!htbp] \centering
  \includegraphics[width=0.6\linewidth]{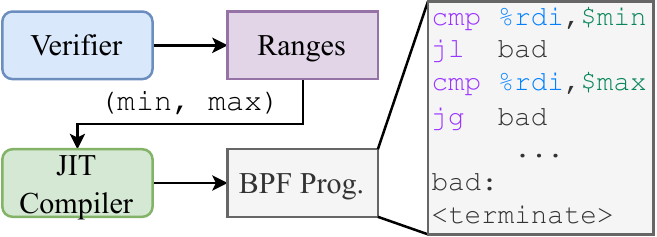} 
  \vspace{-3pt}
  \caption{\mr{DPA Check Generation.}}
\label{fig:dpa_jit} 
\vspace{-10pt} 
\end{figure}

\parh{DPA Check Generation.}~\mr{To deploy DPA from \S~\ref{subsec:dpa}, we
  modify the \ac{BPF} JIT compiler~(\texttt{bpf\_jit\_comp.c} with about 2500 
  LoC) to instrument
  BPF programs. As shown in \F~\ref{fig:dpa_jit}, our modified JIT compiler
  receives a set of expected ranges from the verifier. Then, for each
  parameter, the JIT compiler emits assembly instructions to check whether the
  parameter is within the expected range. If not, we terminate the 
  program~(\texttt{bad} label in \F~\ref{fig:dpa_jit}). This prevents
malicious programs from passing invalid parameters to abuse BPF helpers.}

	\section{Evaluation}
\label{sec:eval}

To evaluate \tool, we first analyze how \tool\ mitigates various attack
interfaces and then benchmark its CVE detectability in
\S~\ref{subsec:sec_eval}. We then assess the performance of \tool under
different \ac{BPF} program setups in \S~\ref{subsec:perf_eval}.

\subsection{Security Evaluation}
\label{subsec:sec_eval}
\subsubsection{Analysis of Attack Mitigation}

\begin{figure}[htbp]
		\vspace{-4pt}
	\centering
	\includegraphics[width=.75\linewidth]{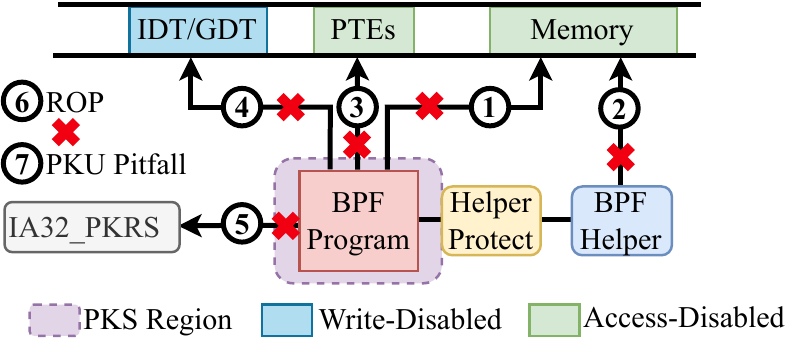}
		\vspace{-4pt}
	\caption{Analysis of attack mitigation.}
	\label{fig:sec_analy}
		\vspace{-10pt}
\end{figure}

We systematically analyze how \tool\ mitigates the representative attacks on the \ac{BPF} ecosystem as well as the potential threats to \tool itself. 
Our analyses are illustrated in \F~\ref{fig:sec_analy}.

\parh{\circled{1} Arbitrary Kernel Access.}~Currently, the most prevalent
threat to the BPF ecosystem is the ability of malicious BPF programs to
arbitrarily modify kernel memory. In order to accomplish this, these \ac{BPF}
programs typically employ corner-case operations to deceive the verifier during
the loading phase and to behave maliciously during runtime. This type of attack
is effectively mitigated due to the fact that \tool\ derives the necessary
memory regions of each \ac{BPF} program and uses \ac{PKS} to prevent any
runtime access beyond this region (\S~\ref{subsec:scalability}), mitigating
such illegal access.

\parh{\circled{2} Helper Function Abuse.}~Apart from launching an attack directly
from \ac{BPF} programs, a malicious \ac{BPF} program may carefully prepare
parameter values 
and pass them to abuse certain helpers. To prevent such abuse, \tool deploys
security enforcement schemes (\S~\ref{subsec:extensibility}) to dynamically
audit helper parameters and also protect critical kernel kernel objects during
the execution of these helpers. Thus, the attacker can no longer take advantage
of these helpers.

\parh{\circled{3} PTE Corruption.}~A page's \ac{PKS} region is configured via
its \ac{PTE}. Consequently, a malicious \ac{BPF} program may attempt to tamper
with the PTEs to disable \tool. However, this is impossible since \tool sets these
PTEs as access-disabled; they are thus protected by \ac{PKS} like other kernel
resources.

\parh{\circled{4} Descriptor Table Tampering.}~Descriptor tables like \ac{GDT}
and \ac{IDT} are essential for segmentation and interrupt handling. Therefore,
blindly setting them as access-disabled would cause system crashes. However,
since these descriptor tables are only accessed in a read-only manner, \tool
sets them as write-disabled, thus preventing malicious \ac{BPF} programs from using them to compromise the kernel.

\parh{\circled{5} Hardware Configuration Tampering.}~Besides memory-based
attacks, attackers may also directly disable \ac{PKS} through
hardware configurations. As described in \S~\ref{subsec:bg-hardware}, \verb*|IA32_PKRS| is a critical register for
configuring \ac{PKS}. One may disable \ac{PKS} by modifying \texttt{IA32\_PKRS}. However, this register can only be modified via special instructions, and \ac{BPF} instruction sets do not include any of these.
Thus, a \ac{BPF} program with these instructions is rejected
immediately. Since the \ac{BPF} programs are set to $W \oplus X$ (meaning write
and executable permissions are not simultaneously enabled), adding these
instructions via self-modification is also impossible.

\parh{\circled{6} Return-Oriented Programming.}~Two properties of
  the \ac{BPF} instruction set prevent potential control-flow hijacking attacks like
  return-oriented programming (ROP). First, \ac{BPF} only supports jump
  instructions with \textit{constant and instruction-level} offsets. This means
  the destinations of jumps are trivially known during the compile time, and
  there are no \textit{unintended ROP gadgets} (jumps between instructions) like
  x86 \cite{rop}. Secondly, as a specialized instruction set, \ac{BPF} does not include any
  instructions that may modify hardware configurations such as \texttt{XRSTOR}
  and \texttt{WRMSR}. These two properties allow \tool to reliably
  detect invalid instructions and prevent \ac{BPF} programs from
tampering with hardware settings.

\parh{\circled{7} Attacks in PKU Pitfalls.}~We carefully examined
  attacks mentioned in PKU Pitfalls~\cite{pitfall}, which focus on breaking
  \ac{PKU}, the user variant of \ac{MPK}. Their noted attacks can be roughly
  categorized into three types. The first type manipulates memory mappings through certain system calls~(e.g., \texttt{mremap}) to
  subvert \ac{PKU}, such as modifying user-space \acp{PTE} and creating mutable backup.
  However, \ac{BPF} programs are incapable of launching such attacks, as there 
  is no helper that can manipulate kernel memory mappings.
  The second type involves tampering with the saved state of
  \texttt{PKRU} and disabling \ac{PKU} entirely upon restoration. Unlike
  \ac{PKU}, \tool exclusively manages the saved state of \texttt{IA32\_PKRS},
  making these attacks infeasible. The third type relies on mechanisms that are
  exclusive to the user space~(e.g., using \texttt{seccomp} to intercept
system calls) and is not applicable to \tool.

\subsubsection{Real-world CVE Evaluation} 

\mr{We surveyed the \ac{BPF} CVEs in the past ten years. A total of 26 CVEs are
  memory exploits~(Appendix.~\hyperref[app:cve_list]{B}) and thus fall within
  the scope of \tool. We tested \tool's effectiveness on all of these
  CVEs. For CVEs with publicly available PoC, we ported and ran the
  PoC on \tool-enabled kernel. For CVEs without PoC, we studied the fixes and
ensure that \tool mitigates them. In sum, we report that \tool successfully
mitigates \textit{all} of them. We now present the following case studies.}

\parh{CVE Case Study.}~To better explain how \tool mitigates these CVEs, we
elaborate on the exploit paths for three of them, 2022-23222, 2020-27194, and 2021-34866.

\noindent \underline{CVE-2022-23222} is a pointer mischeck vulnerability
introduced via a rather new \ac{BPF} feature, \verb*|bpf_ringbuf|. This
new feature was brought to \ac{BPF} in 2020, along with a new pointer type named
\verb*|PTR_TO_MEM_OR_NULL|. However, the verifier had not been updated to track
the bounds of this new type, resulting in this vulnerability. As shown in
\mbox{\F~\ref{fig:cve23222}}, the malicious payload first retrieves a
\verb*|nullptr| via \verb*|ringbuf_reserve|~(line 1), which returns this
newly added pointer type named \verb*|PTR_TO_MEM_OR_NULL|. Since this new type
is not tracked by the verifier, the payload can bypass pointer checks by
tricking the verifier that \texttt{r1} is \texttt{0x0} when it is
\texttt{0x1}~(line 3). \texttt{r1} can then be multiplied with any offset to
perform arbitrary kernel access~(line 6). However, such access
violates \ac{PKS} and is terminated by \tool~(line 7).

\noindent \underline{CVE-2020-27194} is a vulnerability due to incorrect
truncation. As in \F~\ref{fig:cve27194}, the user first inputs an arbitrary
value in the range of \texttt{[0,0x600000001]}~(line 1). Then, the conditional
clause helps the verifier to determine its value range~(line 3). However, when tracking the \verb*|BPF_OR| operator, the
verifier performs a wrong truncation on its upper bound. After the truncation,
the user-controlled \texttt{r5} is viewed by the verifier as a legitimate
constant \texttt{0x1}~(line 5), which is later used as the offset to
perform arbitrary access to the kernel (line 6). Similarly, such access is
stopped by \tool.

\noindent \underline{CVE-2021-34866} is a helper-abuse vulnerability. As shown
in \F~\ref{fig:cve34866}, the malicious payload tries to pass an invalid map to
the \texttt{ringbuf\_reserve} to cause heap overflow~(line 3). However, since
the runtime value of \texttt{r5} does not match the argument of
\texttt{ringbuf\_reserve}, DPA prevents such a mismatched helper call~(line 2). Moreover, supposing that the DPA is not enabled, and the helper tries to tamper
with exploitable kernel objects~(e.g., \texttt{array\_map\_ops}). COP protects 
these objects and thus prevents the helper from accessing them~(line 3). 
Lastly, even if neither COP nor DPA is enabled, and the abused helper manages to
return a leaked kernel pointer, accessing the leaked pointer violates PKS, and
the malicious program is terminated by \tool~(line 4).

\begin{figure}[htbp]
\begin{subcaptionblock}{1\linewidth}
\captionsetup{skip=2pt}
\begin{lstlisting}[language=c]
r0 = ringbuf_reserve(fd, INT_MAX, 0)
r1 = r0 + 1					  // R:r0=0;r1=1 V:r0=r1=?
if (r0 != nullptr)    // R:r0=0;r1=1 V:r0=r1=?
 exit(1)
off = <bad off>     // R:r0=0;r1=1 V:r0=r1=0
off = off * r1		  // R:off=<bad off> V:off=0
*(ptr+off) = 0xbad	// PKS violation
\end{lstlisting}
\caption{Code snippet of CVE-2022-23222}
\label{fig:cve23222}
\smallskip
\begin{lstlisting}[language=c]
r5 = <bad addr>
r6 = 0x600000002
if (r5>=r6||r5<=0) // R&V:0x1<=r5<=0x600000001
 exit(1)
r5 = r5 | 0		   // R:r5=<bad addr> V: r5=0x1
*(ptr+r5)=0xbad  // PKS violation
\end{lstlisting}
\caption{Code snippet of CVE-2020-27194}
\label{fig:cve27194}
\smallskip
\begin{lstlisting}[language=c]
r5 = <bad map fd> 
<DPA checks>                    // DPA violation
r0=ringbuf_reserve(r5,INT_MAX,0)// COP-guarded
*(r0+ptr_off) = 0xbad           // PKS violation
\end{lstlisting}
\caption{Code snippet of CVE-2021-34866}
\label{fig:cve34866}
\end{subcaptionblock}
\caption{CVE case study. \texttt{R} denotes variable runtime statuses. \texttt{V} denotes verifier-deduced values of variables.}

\end{figure}

\subsection{Performance Evaluation}
\label{subsec:perf_eval}

\parh{Evaluation Setup.}~We assess \tool\ performance on Linux v6.1.38 and a
5-core Intel 8505 processor with PKS support. \mr{To reduce variance,
  hyper-threading, turbo-boost, and frequency scaling are disabled.} \fixme{All
    evaluated \ac{BPF} programs are executed in the \ac{JIT} mode, given that
  \ac{BPF} \ac{JIT} is enabled by default on all supported platforms.}
  Moreover, both COP and DPA (\fixme{\S~\ref{subsec:extensibility}}) are
  enabled; COP is configured to protect the critical objects identified in
  \S~\ref{subsec:cop}. \mr{We manually inspected the CPU utilization to ensure 
  it is close to 100\%, and that the overhead is not hidden by the increased CPU load.}

\subsubsection{Micro Benchmark}
\label{subsec:micro_bench}

For the micro benchmark, we measure the CPU cycles of four key operations in \tool.
We list the four operations in \T~\ref{table:micro_bench}. \texttt{set\_pkrs}
changes region permissions  by changing \texttt{IA32\_PKRS} via \texttt{WRMSR}.
\texttt{get\_pkrs} returns the current permission configuration by reading
\texttt{IA32\_PKRS} via \texttt{RDMSR}. \texttt{bpf\_\{entry/exit\}} is the
total cost of entering/exiting a \ac{BPF} program, which includes operations like
swapping stack, managing \ac{BPF} context, and configuring region permissions with
\texttt{set\_pkrs}. \texttt{dpa\_check\_args} is the cost of checking helper
parameters. Each operation is measured by averaging ten runs of one million
invocations to eliminate randomness. Since Intel has introduced the concept of
``performance core'' and ``efficient core'', we measure their cycles
independently.

As shown in \T~\ref{table:micro_bench}, the overall switching cost of \tool is
less than 200 cycles, which is negligible for most BPF programs~(see
\S~\ref{subsec:macro_bench} for details). Notably, setting and getting the
region permissions~(\texttt{set\_pkrs}/\texttt{get\_pkrs}) in \ac{PKS} is more
expensive than its user-space variant in \verb*|libmpk|~\cite{libmpk}
(see the caption of \T~\ref{table:micro_bench}). We presume that this
is because, in \ac{PKU}, the region permission is controlled via a
dedicated register named \texttt{PKRU} with two special instructions
\texttt{RDPKRU/WRPKRU}, whereas in \ac{PKS} employed by \tool, its region
permission is stored in an \ac{MSR} named \texttt{IA32\_PKRS} without any
special instruction. To configure the permission in \texttt{IA32\_PKRS}, one
has to use the \texttt{RDMSR/WRMSR} instructions with the \ac{MSR} ID
\texttt{0x6E1}. Moreover, although the cost of \texttt{dpa\_check\_args}
varies based on the checked range type~(e.g., value point \texttt{[0x1,0x1]} costs
less than value range \texttt{[0x1,0x10]}), we report that these costs are all
less than ten cycles. Lastly, we observe that the operations of \tool are not
substantially affected by the difference between performance and efficient cores.

\begin{table}[htbp]
	\centering
	\caption{Micro benchmark results. We use P to denote the cycles of performance cores and E for the efficient cores. As a reference~\cite{libmpk}, user-space \texttt{RDPKRU},
		\texttt{WRPKRU} take 0.5 and 23.3 cycles, respectively.}
	\label{table:micro_bench}
	\resizebox{0.8\linewidth}{!}{
		\begin{tabular}{llll}
			\hline
			\textbf{Operation}       	& \multicolumn{2}{l}{\textbf{\#Cycles}} & \textbf{Note} \\ \hline
			\texttt{get\_pkrs/RDMSR}    & P:36  & E:43   & Get region permissions \\ \hline
			\texttt{set\_pkrs/WRMSR}    & P:111 & E:112  & Set region permissions \\ \hline
			\texttt{bpf\_\{entry/exit\}}& P:154 & E:173  & Entry/exit BPF program \\ \hline
			\texttt{dpa\_check\_args}   & P:$\leq$10& E:$\leq$10    & Check helper arguments \\ \hline
		\end{tabular}
	}
\end{table}

\parh{\tool's Overhead vs. \#BPF Programs.}~To show \tool supports that numerous
\ac{BPF} programs, we prepare the following experiments. We attach 1, 10, 32,
64, and 128 \ac{BPF} programs to trace \texttt{execve}\footnote{We clarify that one
\ac{BPF} tracepoint only supports up to 64 programs~\cite{tracelimit}, so we
attach to both entry and exit tracepoints of \texttt{execve} in these
experiments.}, run a program that continuously creates processes 
for one minute, and measure the number of processes created. In this setting,
each invocation of \texttt{execve} stresses \tool to constantly switch between the BPF programs. \fixme{Moreover, we
craft each \ac{BPF} program as succinct (programs that directly return) so that
\tool's relative overhead is not ``dominated'' by the overly lengthy \ac{BPF}
programs.}

\begin{figure}[!htbp]
	\centering
	\includegraphics[width=.9\linewidth]{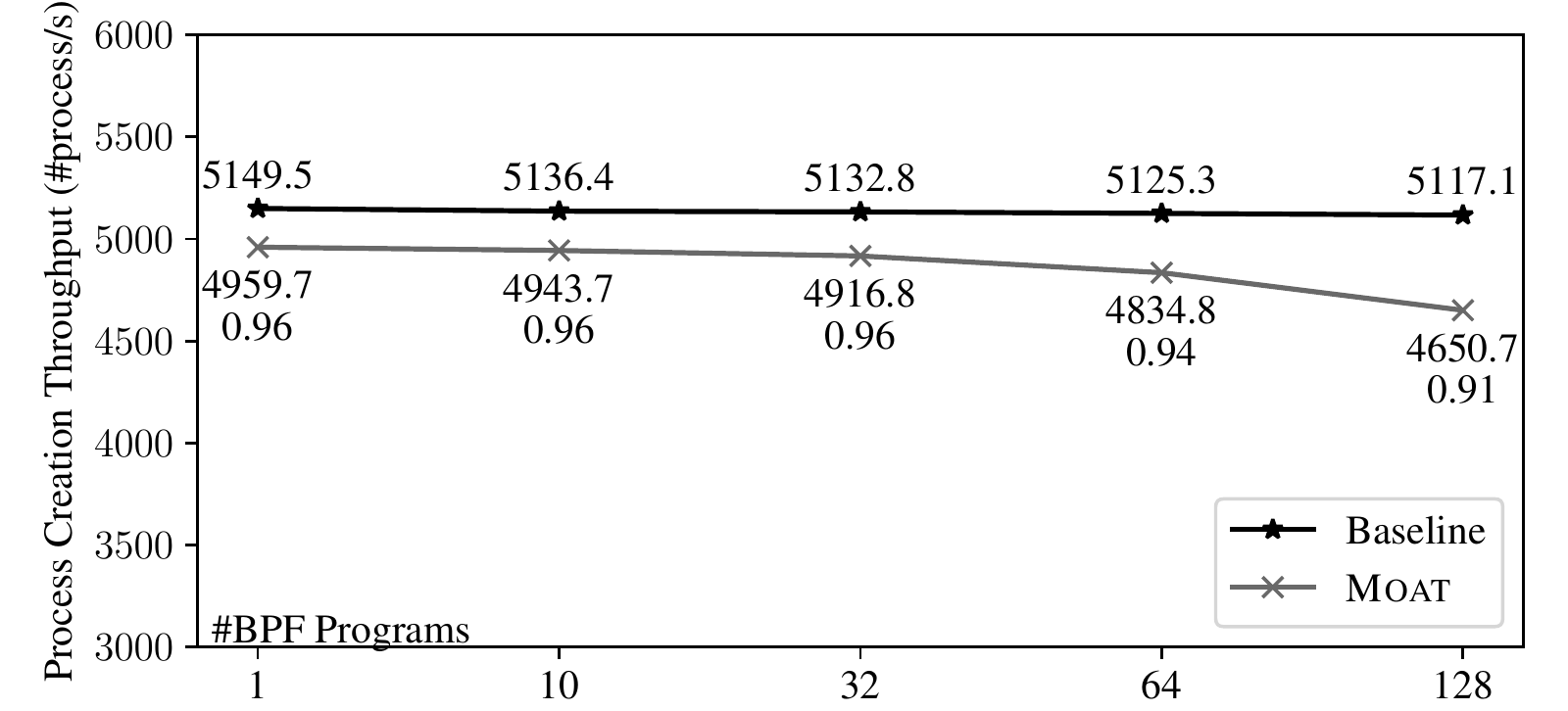}
	\caption{\tool's overhead with respect to \#BPF programs.}
	\label{fig:syscall_eval}
\end{figure}

We report our results in \F~\ref{fig:syscall_eval}. Since we use simple \ac{BPF}
programs, the baseline performance (without \tool) is not observably affected by
the number of \ac{BPF} programs; we expect that in real-world cases, the baseline
performance would also drop as the number of \ac{BPF} programs increases.
\tool's overhead stays largely the same~(4\%) when there are 1, 10, or 32 \ac{BPF}
programs and then slowly increases with the number of \ac{BPF} programs. It
eventually incurs 9\% overhead with 128 \ac{BPF} programs. With further
inspection, we find that over 64 \ac{BPF} programs with isolated TLB entries
pose heavy stress to TLB, resulting in increased overhead. However, having
over 64 \ac{BPF} programs attached to the same place is extremely rare~(if
it occurs at all). Nonetheless, even in such cases, \tool incurs a performance penalty of less than 10\%.

To complement the above experiment, we prepare the following experiment
where \ac{BPF} programs are attached to different kernel locations. There are, in
total, 685 \ac{BPF} tracepoints of system calls in Linux~\cite{tracelist}.
Following a similar setting as above, we attach each of these tracepoints with a
simple \ac{BPF} program and run UnixBench~\cite{unixbench} to measure the
overall system performance. In sum, there are nearly 700 \ac{BPF} programs in
the kernel with diverse invocation patterns. Thus, this setting stresses the
system in a manner distinct from the previous experiment. \fixme{We report that in such
settings, the average UnixBench baseline score is 4936.3, \tool's score is
4720.8, and thus, the incurred overhead is about 5\%.}

\fixme{From these two experiments, we interpret that the overhead of
  \tool is slightly affected by the number of \ac{BPF} programs due to the TLB stress. Nevertheless, \tool's overhead falls in a reasonable and
  promising range (4\%\textasciitilde9\%), even for cases that are much heavier
  and rarely seen in real-world ones.}

\subsubsection{Macro Benchmark} \label{subsec:macro_bench} For the macro
benchmark, we set up three mainstream \ac{BPF} use cases: network, system
tracing, and system-call filtering. \mr{On the network cases~(i.e., Socket and
  XDP), we use the smallest applicable packet size because \ac{BPF} operates on
each packet. Thus, under the same bandwidth, smaller-sized packets will incur higher
throughput (i.e., more packets) and lead to more invocations of BPF 
programs, eventually putting 
more stress on \tool.}

\parh{Network --- Socket.} To evaluate \tool's overhead on the network
applications, we simulate a traffic monitoring scenario. A traffic generator
sends UDP packets for one minute, with a packet size of 16 bytes, to our tested
device. A server on the tested device receives these packets. Both the sender
and receiver have a 1\,GbE network interface controller. As for the tested
\ac{BPF} programs, we use five socket filtering programs from the Linux source
tree, similar to previous works~\cite{evil}:

\begin{itemize}[leftmargin=.35cm,parsep=0pt,topsep=0pt]
\item \texttt{drop}: directly ignores the packet.
\item \texttt{byte}: monitors the traffic in bytes from each protocol.
\item \texttt{pkt}: monitors the traffic in packets from each protocol.
\item \texttt{trim}: only keeps the packet header to the socket.
\item \texttt{flow}: monitors the network traffic based on protocol, interface,
  source, destination, and port.
\end{itemize}

We attach a socket to the receiver's network interface and set up these
\ac{BPF} programs to monitor the network traffic over the socket. In addition
to the evaluation of each program, we also conduct a full-on
experiment where five \ac{BPF} programs are attached simultaneously to stress \tool.

\begin{table}[htbp]
	\caption{\tool's traffic monitoring performance in Thousand Packets per
	Second~(TPPS). The full-on experiment is denoted as ``all''. The ``vanilla''
	throughput without \ac{BPF} program is 596.3 TPPS; the relative throughput
	is denoted in parentheses, e.g., (99.73\%).}
	\label{table:skfilter_eval}
	\resizebox{.98\linewidth}{!}{
	\begin{tabular}{ccccccc}
		\hline
		\multicolumn{1}{c}{\textbf{Throughput (TPPS)}} & \textbf{drop}                          & \textbf{byte}                                              & \textbf{pkt}                                               & \textbf{trim}                                              & \textbf{flow}                                              & \textbf{all}                                               \\ \hline
		\textbf{Baseline}                                  & \begin{tabular}[c]{@{}c@{}}594.39\\ (99.70\%)\end{tabular} & \begin{tabular}[c]{@{}c@{}}594.67\\ (99.73\%)\end{tabular} & \begin{tabular}[c]{@{}c@{}}594.26\\ (99.66\%)\end{tabular} & \begin{tabular}[c]{@{}c@{}}594.74\\ (99.73\%)\end{tabular} & \begin{tabular}[c]{@{}c@{}}594.39\\ (99.68\%)\end{tabular} & \begin{tabular}[c]{@{}c@{}}587.22\\ (98.47\%)\end{tabular} \\ \hline
		\textbf{\tool}                                  & \begin{tabular}[c]{@{}c@{}}593.10\\ (99.46\%)\end{tabular} & \begin{tabular}[c]{@{}c@{}}594.31\\ (99.66\%)\end{tabular} & \begin{tabular}[c]{@{}c@{}}594.43\\ (99.68\%)\end{tabular} & \begin{tabular}[c]{@{}c@{}}594.69\\ (99.73\%)\end{tabular} & \begin{tabular}[c]{@{}c@{}}593.10\\ (99.46\%)\end{tabular} & \begin{tabular}[c]{@{}c@{}}575.33\\ (96.48\%)\end{tabular} \\ \hline
	\end{tabular}
}
\end{table}
\fixme{As shown in \T~\ref{table:skfilter_eval}, \tool incurs negligible overhead~(<1\%) for all \ac{BPF} programs if they are solely executing in the kernel. Even in the full-on experiment, which forces \tool to constantly switch between these \ac{BPF} programs, \tool brings only a very small throughput drop of 2\%.}

\parh{Network --- XDP.} Besides processing packets from a socket buffer,
\ac{BPF} provides a direct way to control the network --- eXpress Data
Path~(XDP). XDP processes packets at an early stage in the network stack to achieve fast packet processing. Following the
settings in the socket experiment, we simulate a packet processing scenario.
Similar to prior works~\cite{evil}, we run five XDP programs from the Linux
source tree:

\begin{itemize}[leftmargin=.35cm,parsep=0pt,topsep=0pt]
	\item \texttt{xdp1}: parses the IP header, keeps packets count in a \ac{BPF} map, and drops the packets.
	\item \texttt{xdp2}: same as \texttt{xdp1}, but re-sends the packets.
	\item \texttt{adj}: trims the packets into ICMP packets, sends them back,
	and keeps packet count in a \ac{BPF} map.
	\item \texttt{rxq1}: counts and drops the packets in each receive queue.
	\item \texttt{rxq2}: same as \texttt{rxq1}, but re-sends the packets.
\end{itemize}

Unlike socket filters, XDP programs require packets to be over a certain
size, so we tune our traffic generator to send packets that go through
the maximum possible execution path of the tested programs.
We send packets of 64 bytes for \texttt{xdp1} and \texttt{xdp2} and packets of
100 bytes for \texttt{adj}, \texttt{rxq1}, and \texttt{rxq2}.

\begin{table}[htbp]
	\caption{\tool's XDP performance in TPPS. The ``vanilla'' throughput without XDP program is 532.9 TPPS with 100-byte packets, and 561.5 TPPS with 64-byte packets; the relative throughput is denoted in parentheses, e.g.,~(99.55\%).}
	\label{table:xdp_eval}
	\centering
	\resizebox{.98\linewidth}{!}{
	\begin{tabular}{cccccc}
		\hline
		\multicolumn{1}{c}{\textbf{Throughput (TPPS)}} & \textbf{xdp1}                                              & \textbf{xdp2}                                              & \textbf{adj}                                               & \textbf{rxq1}                                              & \textbf{rxq2}                                              \\ \hline
		\textbf{Baseline}                                  & \begin{tabular}[c]{@{}c@{}}560.58\\ (99.84\%)\end{tabular} & \begin{tabular}[c]{@{}c@{}}557.78\\ (99.34\%)\end{tabular} & \begin{tabular}[c]{@{}c@{}}531.11\\ (99.66\%)\end{tabular} & \begin{tabular}[c]{@{}c@{}}528.36\\ (99.15\%)\end{tabular} & \begin{tabular}[c]{@{}c@{}}530.52\\ (99.55\%)\end{tabular} \\ \hline
		\textbf{\tool}                                  & \begin{tabular}[c]{@{}c@{}}560.15\\ (99.76\%)\end{tabular} & \begin{tabular}[c]{@{}c@{}}557.76\\ (99.33\%)\end{tabular} & \begin{tabular}[c]{@{}c@{}}530.65\\ (99.58\%)\end{tabular} & \begin{tabular}[c]{@{}c@{}}527.57\\ (99.00\%)\end{tabular} & \begin{tabular}[c]{@{}c@{}}527.66\\ (99.05\%)\end{tabular} \\ \hline
	\end{tabular}
}
\end{table}

As illustrated in \T~\ref{table:xdp_eval}, \tool incurs negligible performance
penalties~(<1\%) when executing XDP programs.

\parh{System Tracing.} System tracing is another mainstream \ac{BPF} use case.
To evaluate \tool's overhead on system tracing, we prepare 11 \ac{BPF} programs
to trace frequent system events like page faults, process creation, context
switch, and file operations. These programs collect relevant system statistics
for user-space analysis. Then, we run UnixBench~\cite{unixbench} to measure
the overall system performance. UnixBench includes the following tests:
\ding{192} execl throughput, \ding{193} file copy, \ding{194} pipe throughput,
\ding{195} pipe-based context switching, \ding{196} process creation,
\ding{197} shell scripts, and \ding{198} system call.

\begin{figure}[htbp]
	\centering
	\includegraphics[width=.85\linewidth]{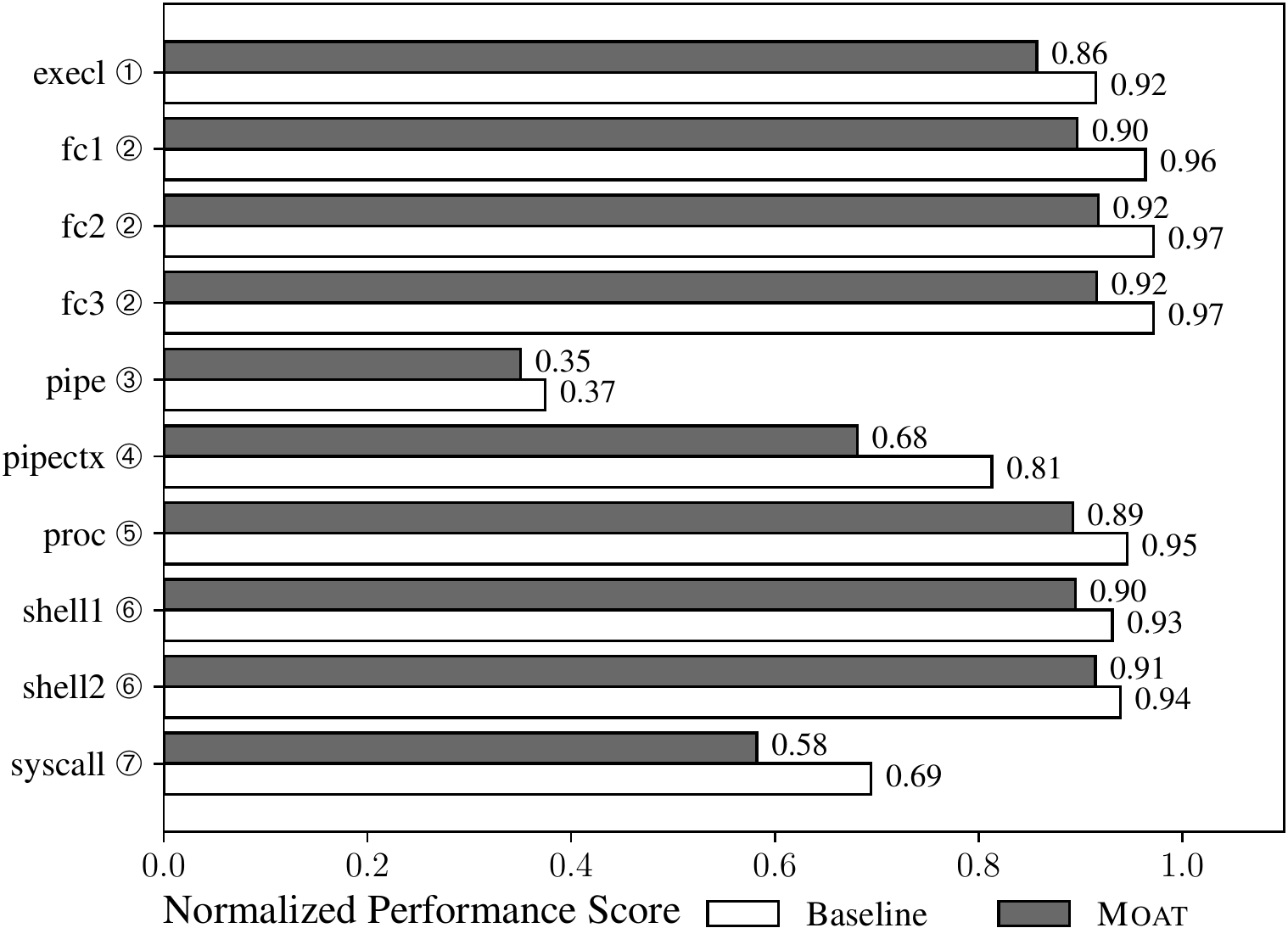}
	\caption{UnixBench normalized scores with respect to the ``vanilla'' scores without \ac{BPF} tracepoints. The 
	``fc* \ding{193}'' and \mbox{``shell* \ding{197}''} are file copy tests and shell tests with different settings.}
	\label{fig:tp_eval}
\end{figure}

We report the results in \F~\ref{fig:tp_eval}. We find that \tool imposes a
small slowdown~($\leq$\,6\%) for most UnixBench tests. The maximum performance
loss brought by \tool is 13\% in test~\ding{198}. Such overhead seems moderate.
However, note that the BPF programs without \tool already bring a non-trivial
performance penalty~(e.g., 63\% slowdown in test \ding{194}). Therefore, the
performance loss brought by \tool~(<13\%) is reasonably low for system tracing.

\parh{Syscall Filtering.} \ac{BPF} is also used to enhance software
security~\cite{seccomp,sysfilter,ringguard}. \texttt{seccomp-BPF} allows
filtering the system calls of a process with \ac{BPF}.
\texttt{sysfilter}~\cite{sysfilter} is an automated tool that analyzes a
program, creates the set of system calls the program needs, and restricts the
program using \texttt{seccomp-BPF}. We use \texttt{seccomp-BPF} and
\texttt{sysfilter} to evaluate \tool's overhead in such a use case. We apply
the \tool-hardened filter to Nginx and benchmark it using
\texttt{wrk}~\cite{wrk} with one, two, and three client processes; each sends
requests for one minute with 128 connections. Nginx is configured with the same
number of worker processes. \mr{To clarify this setting: the number of 
processes for Nginx
  typically shall not exceed the number of cores, and adding more processes
  does not increase the throughput due to the context-switch cost.} All requests
  are sent over the loopback~(\texttt{lo}) to minimize network
  interference.

\begin{table}[htbp]
	\caption{Nginx throughput in Thousands of Request per Second~(Treq/s). The relative throughput is in the parentheses, e.g.,~(95.6\%).}
	\label{table:nginx_eval}
	\centering
	\resizebox{.85\linewidth}{!}{
\begin{tabular}{clll}
	\hline
	\textbf{Throughout (Treq/s)}                                                & \multicolumn{1}{c}{\textbf{1 worker}}                         & \multicolumn{1}{c}{\textbf{2 worker}}                         & \multicolumn{1}{c}{\textbf{3 worker}}                         \\ \hline
	\begin{tabular}[c]{@{}c@{}}\textbf{Vanilla}\\ (no \texttt{seccomp-BPF})\end{tabular} & \begin{tabular}[c]{@{}l@{}}148.1 (100\%)\\ $\pm$12.81\end{tabular} & \begin{tabular}[c]{@{}l@{}}179.5 (100\%)\\$\pm$ 8.35\end{tabular}  & \begin{tabular}[c]{@{}l@{}}165.2 (100\%)\\ $\pm$4.72\end{tabular}  \\ \hline
	\textbf{Baseline}                                                           & \begin{tabular}[c]{@{}l@{}}147.2 (99.4\%)\\ $\pm$9.56\end{tabular} & \begin{tabular}[c]{@{}l@{}}171.3 (95.4\%)\\ $\pm$8.08\end{tabular} & \begin{tabular}[c]{@{}l@{}}160.5 (97.2\%)\\ $\pm$5.28\end{tabular} \\ \hline
	\textbf{\tool}                                                               & \begin{tabular}[c]{@{}l@{}}142.3 (96.1\%)\\ $\pm$8.77\end{tabular} & \begin{tabular}[c]{@{}l@{}}166.3 (92.6\%)\\ $\pm$6.70\end{tabular} & \begin{tabular}[c]{@{}l@{}}158.0 (95.6\%)\\ $\pm$4.48\end{tabular} \\ \hline
\end{tabular}
}
\vspace{-10pt}
\end{table}

As shown in \T~\ref{table:nginx_eval}, \tool incurs an additional throughput
drop of 3\%. Moreover, the standard deviations of throughput are within
the normal range. Therefore, \tool does not introduce fluctuation to Nginx
throughput.

\subsection{Additional Evaluation}
\label{subsec:add_eval}

In addition to the above experiments, we also evaluate \tool's memory
footprints, instrumentation cost from DPA, and compare \tool's performance with
a prior work~\cite{sandbpf}.

\mr{ \parh{Memory Footprint.}~As mentioned in \S~\ref{sec:implementation},
  \tool aligns BPF-related objects~(e.g., maps) to 4\,KB to ensure that they do
  not interleave with other kernel structures, introducing extra memory
  footprints.  We provide a detailed breakdown of \tool's memory footprints in
  \T~\ref{table:mem_footprint}. Specifically, \tool uses four pages to set up
  the page table of isolated address-spaces, one page for the stack and one for
  the context. As for critical objects identified in \S~\ref{subsec:cop}, \tool
  uses an extra page to toggle their permissions independently.  Additionally,
  the memory used by the BPF program binaries and BPF maps is aligned up to a 
  multiple of page size.

\begin{table}[htbp]
\centering
\caption{\mr{Breakdown of \tool's memory footprint. AS: Address Space; ST: 
Stack; Ctx: Context; CO: Critical Objects. $P$ and $M$ denote \#pages of 
program and map, respectively.}}
\label{table:mem_footprint}
\resizebox{.65\linewidth}{!}{
  \begin{tabular}{ccccccc}
  \hline
  \textbf{Type}   & AS & ST & CO & Ctx & Prog                  & Map \\ \hline
  \textbf{\#Page} & 4  & 1  & 1  & 1   & $\lceil P \rceil$     & $\lceil M 
  \rceil$  \\ \hline
  \end{tabular}
}
\vspace{-15pt}
\end{table}

Though \tool's memory footprints seem non-trivial from \T~\ref{table:mem_footprint},
we clarify that they are static and thus do not grow dynamically during the
runtime. Even if there are thousands of BPF programs, \tool's memory footprints
are merely a few megabytes, which is negligible for modern systems.

\parh{Helper Instrumentation Cost.}~As described in \S~\ref{subsec:dpa}, \tool
instruments the BPF programs to insert DPA checks.
\T~\ref{table:instr} shows the number of helper calls and memory access
made by the BPF programs\footnote{The BPF programs in syscall filtering
scenario are cBPF programs and thus do not support helper functions; we do not
include them here.} in \S~\ref{subsec:macro_bench}. The former reflects the 
number of 
DPA
checks inserted by \tool, while the latter reflects the number of checks
from SFI-based solutions. We report that, in most cases, the number of DPA
checks is smaller than that of SFI-based solutions, let alone that SFI only 
offers memory safety, which is offered by PKS in \tool.

\begin{table}[htbp] \centering \caption{\mr{The number of checks of DPA and
  SFI-based isolation. Note that the tracepoint~(marked with *) consists
multiple BPF programs, and we report their average number of inserted checks.}}
\label{table:instr} 
\setlength{\tabcolsep}{2pt}
\resizebox{.9\linewidth}{!}{ \begin{tabular}{lcccccccccc}
    \hline \textbf{Name} & \textbf{drop} & \textbf{byte} & \textbf{pkt} &
    \textbf{trim} & \textbf{flow} & \textbf{xdp1} & \textbf{xdp2} &
    \textbf{adj} & \textbf{rxq} & \textbf{tracepoint*} \\ \hline
\textbf{\#Helper} & 0 & 1 & 1 & 0 & 2 & 3 & 3 & 5 & 3  & 6.4 \\ \hline
\textbf{\#Mem.} & 0 & 4 & 3 & 1 & 61 & 16 & 29 & 53 & 37 & 6.9 \\ \hline
\end{tabular} } 
\vspace{-10pt}
\end{table}

\parh{Comparison with SandBPF.}~As far as we know, SandBPF~\cite{sandbpf} is
the only work on BPF isolation at the time of writing. In this section, we 
compare \tool with
SandBPF. Technically, SandBPF enforces isolation by inserting
software runtime checks into the memory access of \ac{BPF} programs. As a
result, it shall incur a relatively higher overhead than hardware-based
methods. To substantiate this observation, we conduct the following direct
comparative study. The authors of SandBPF conduct an evaluation with Phoronix 
Test
Suite~\cite{pts}; we reproduce the same experiments on \tool. Note that the 
source code of SandBPF is not public at the time of writing, so we directly 
refer to their data in \T~\ref{table:sandbpf}. 

\begin{table}[htbp]
\centering
\caption{\mr{Comparison with SandBPF~\cite{sandbpf}. We provide \tool's 
relative overhead~(Rel.) and SandBPF's overhead~(Ref.).}}
\label{table:sandbpf}
\setlength{\tabcolsep}{4pt}
\resizebox{.85\columnwidth}{!}{%
\begin{tabular}{lcccc|cccc}
\hline
\multicolumn{1}{c}{\multirow{2}{*}{\textbf{\begin{tabular}[c]{@{}c@{}}Test \#Conn\\ (req./s)\end{tabular}}}} &
  \multicolumn{4}{c|}{\textbf{XDP}} &
  \multicolumn{4}{c}{\textbf{Socket Filter}} \\ \cline{2-9} 
  \multicolumn{1}{c}{} & \textbf{Base}  & \textbf{\tool}  & \textbf{Rel.}       
  & 
  \textbf{Ref.}           
  & \textbf{Base}   & 
  \textbf{\tool}  & \textbf{Rel.}   & \textbf{Ref.}               \\ \hline
\textbf{Apache 20}   & 34,303 & 33,689 & 2\%        & \textbf{0\%}   & 40,666 & 
40,286 & \textbf{1\%} & 4\%          \\ \hline
\textbf{Apache 100}  & 31,929 & 30,726 & \textbf{4\%} & 8\%          & 37,998 & 
36,546 & \textbf{4\%} & 4\%          \\ \hline
\textbf{Apache 200}  & 27,751 & 26,657 & \textbf{4\%} & 5\%          & 32,652 & 
31,344 & 4\%          & \textbf{3\%} \\ \hline
\textbf{Apache 500}  & 24,786 & 24,439 & \textbf{1\%} & 7\%          & 30,262 & 
29,423 & \textbf{3\%} & 7\%          \\ \hline
\textbf{Apache 1000} & 24,597 & 24,470 & \textbf{1\%} & 6\%          & 29,545 & 
28,961 & \textbf{2\%} & 7\%          \\ \hline
\textbf{Nginx 20}    & 22,688 & 21,892 & \textbf{3\%} & 7\%          & 23,359 & 
23,530 & \textbf{0\%} & 10\%         \\ \hline
\textbf{Nginx 100}   & 21,492 & 20,689 & \textbf{4\%} & 7\%          & 22,870 & 
22,482 & \textbf{2\%} & 8\%          \\ \hline
\textbf{Nginx 200}   & 19,972 & 19,216 & \textbf{4\%} & 6\%          & 21,562 & 
20,984 & \textbf{3\%} & 8\%          \\ \hline
\textbf{Nginx 500}   & 18,470 & 17,814 & \textbf{4\%} & 6\%          & 19,421 & 
18,713 & \textbf{4\%} & 7\%          \\ \hline
\textbf{Nginx 1000}  & 17,024 & 16,735 & \textbf{2\%} & 3\%          & 17,392 & 
17,098 & \textbf{2\%} & 6\%          \\ \hline
\end{tabular}%
}
\vspace{-10pt}
\end{table}

As shown in \T~\ref{table:sandbpf}, \tool's overhead is lower than SandBPF in
most testcases~(Rel. v.s. Ref.). The \tool's highest overhead is 4\%, while 
SandBPF's is 10\%. Again, we interpret the advantage of \tool is 
attributed to the reduced number of inserted checks compared to SFI-based 
approaches.
}

	\section{Related Work}
\label{sec:rw}

\mr{In this section,
	we discuss other related works on MPK and BPF. We already reviewed highly 
	relevant works on kernel isolation and compared
\tool's design with theirs in \S~\ref{subsec:design_comparison}.}

\parh{MPK-based Isolation.}~Prior to \ac{PKS}, Intel first announced its
user-space variant \ac{PKU}. Consequently, most existing
works~\cite{libmpk,vdom,epk} using \ac{MPK} focus on user-space isolation. To
better utilize \ac{PKU} as an isolation primitive, \citet{libmpk} proposed
\verb*|libmpk|, which resolves the semantic discrepancies between \ac{PKU} and
conventional \verb*|mprotect|. \fixme{VDom and EPK~\cite{vdom,epk} aim to
  provide unlimited keys in the \textit{user space} via key virtualization.
  Despite the similarity, we clarify that isolating \ac{BPF} programs in the
  \textit{kernel} is a distinct scenario and comes with its own challenges.
  This is why we have proposed the lightweight two-layer design to efficiently
isolate \ac{BPF} programs.} Apart from using \ac{PKU} to isolate user
applications, efforts are made to isolate trusted applications in SGX via
\ac{PKU}~\cite{sgxlock,enclavedom}. SGXLock~\cite{sgxlock} establishes mutual
distrust between the kernel and the trusted SGX applications, while
EnclaveDom~\cite{enclavedom} enables intra-isolation within one enclave.
\ac{PKU} has also been used for kernel security. IskiOS~\cite{iskios} applies
\ac{PKU} to kernel pages by marking them as user-owned.

\parh{BPF Security.}~There are prior
works~\cite{formaljit,jitk,prevail,prsafe,exobpf} on securing the BPF
ecosystem. Most of them use formal methods to enhance the following \ac{BPF}
components: the verifier, the JIT compiler, or the BPF program.
To enhance the \ac{BPF} verifier, \citet{prevail} propose PREVAIL based on
abstract interpretation, which supports more program structures (e.g., loops)
and is more efficient than the standard verifier. PRSafe~\cite{prsafe}, on the
other hand, designs a new domain-specific language, whose ultimate goal is to
build a mathematically verifiable compiler for \ac{BPF} programs.
Jitk~\cite{jitk} offers a JIT compiler whose correctness is proven manually,
and \citet{formaljit} generate automated proof for real-world \ac{BPF} JIT
compilers. Lastly, \citet{exobpf} build proof-carrying \ac{BPF} programs,
requiring developers to provide a correctness proof with the program before
loading it into the kernel.

	\section{Limitations}
\label{sec:discussion}

\mr{
In this section, we discuss the limitations of \tool, including the 
unidentified critical objects, issues brought by the granularity of PKS, and 
potential barriers to deploying \tool.

\parh{Unidentified Critical Objects.}~In \S~\ref{subsec:cop}, we manually
identified the critical objects in the BPF subsystem that have been 
exploited in the wild or similar to the ones that have been exploited.
Despite two authors' rich experience, there might still exist
unidentified critical objects. However, to exploit these unidentified objects,
one still has to find a BPF helper that can be abused to access these critical
objects and bypass the parameter checks~(i.e., DPA) enforced by \tool. Thus,
even if few unidentified critical objects exist, they would be hard, if at all 
possible, to exploit due to 
DPA.

\parh{Page-size Granularity.}~\tool leverages PKS to enforce isolation, which
only supports 4\,KB granularity. Though we carefully adjust BPF-related objects
so that they are page-aligned, there still exist few corner-cases where PKS
does not apply. In particular, BPF programs shall only access certain fields
of the context \texttt{sk\_buff}, which is enforced via the static checks by
the verifier. However, applying PKS to these fields would 
significantly bloat \texttt{sk\_buff} due to the granularity. \tool thus cannot 
use PKS to constrain the access of BPF programs. As a result, \texttt{sk\_buff} 
might have some bits leaked 
to BPF programs if the verifier's static checks are bypassed. Fortunately, BPF 
programs only receive a 
copy of
\texttt{sk\_buff} and cannot tamper with the original structure. Therefore, the
consequence of such leakage, per our observation, seems trivial.

\parh{Barrier to Deploying \tool.}~To deploy \tool on other platforms or
systems, there exist several barriers. First, \tool requires a hardware feature
like PKS that can provide \mbox{page-level} isolation. Fortunately, major
platforms already support security features with similar capability. For
example, on RISC-V, a prior work~\cite{donky} implements a PKS-like feature
named Donky, which could be used to support \tool. Therefore, we expect \tool
to be deployable on other platforms with moderate engineering effort. Second, as
we mentioned in \S~\ref{subsec:add_eval}, \tool introduces additional memory
footprints. Though the introduced footprints are only a few pages and
negligible for modern systems, it might still create obstacles for some
embedded systems, where memory is a scarce resource. Third, \tool consists of
about 3,000 lines of C code, which requires moderate engineering effort to port to
other platforms.}

\section{Conclusion}
Despite using \ac{BPF} to extend kernel functionality, malicious BPF
applications can bypass static security checks and conduct unauthorized kernel
accesses. We present \tool\ to isolate potentially malicious BPF applications
from the kernel. \tool\ delivers practical and extensible protection with
a low cost, in compensation to contemporary BPF verifiers. 

	\section*{Acknowledgments}
	We are grateful to the anonymous reviewers, Dr. Zhou Lei, \mbox{Dr. Adrian 
	Rowland}, and the members of COMPASS Lab for their valuable comments.
This work is partly supported by the National Natural Science Foundation of 
China under Grant No.62372218 and Shenzhen Science and Technology Program under 
Grant No.SGDX20201103095408029. The HKUST authors are supported in part by a RGC CRF grant under the contract C6015-23G.
	\bibliographystyle{plainnat}
	\bibliography{usenix.bib}
	
  \section*{Appendix A: Critical Objects} \mr{\label{app:cop} We list the
    critical objects identified by us in
    \T~\ref{table:cotable}.} We categorize the objects in 
    \T~\ref{table:cotable} based 
on their locations. Despite different location~(i.e., map and iterator), 
both of them are used to implement dynamic dispatching in the kernel. 

\begin{table}[H] \caption{Critical objects in the BPF subsystem.}
  \label{table:cotable} \centering \resizebox{.8\linewidth}{!}{
    \begin{tabular}{cl} \hline \textbf{Location} & \textbf{Critical Object} \\
      \hline \multicolumn{1}{c}{\begin{tabular}[c]{@{}c@{}}Map\end{tabular}} & 
      \begin{tabular}[c]{@{}l@{}}
        \texttt{array\_map\_ops}, \texttt{percpu\_array\_map\_ops},\\
        \texttt{prog\_array\_map\_ops}, \texttt{sock\_map\_ops},\\
        \texttt{cgroup\_storage\_map\_ops}, \texttt{htab\_map\_ops},\\
        \texttt{htab\_percpu\_map\_ops}, \texttt{htab\_lru\_map\_ops},\\
        \texttt{htab\_lru\_percpu\_map\_ops}, \texttt{trie\_map\_ops},\\
        \texttt{task\_storage\_map\_ops}, \texttt{dev\_map\_ops},\\
        \texttt{sk\_storage\_map\_ops}, \texttt{cpu\_map\_ops},\\
        \texttt{xsk\_map\_ops}, \texttt{perf\_event\_array\_map\_ops},\\
        \texttt{queue\_map\_ops}, \texttt{stack\_map\_ops},\\
        \texttt{bpf\_struct\_ops\_map\_ops}, \texttt{ringbuf\_map\_ops},\\
        \texttt{bloom\_filter\_map\_ops},\texttt{cgroup\_storage\_map\_ops},\\
        \texttt{cgroup\_array\_map\_ops},\texttt{array\_of\_maps\_map\_ops},\\
        \texttt{stack\_trace\_map\_ops}, \texttt{htab\_of\_maps\_map\_ops}, \\
        \texttt{user\_ringbuf\_map\_ops}, \texttt{inode\_storage\_map\_ops}
        \end{tabular} \\ \hline
        \multicolumn{1}{c}{\begin{tabular}[c]{@{}c@{}}Iterator\end{tabular}} & 
        \begin{tabular}[c]{@{}l@{}}
          \texttt{cgroup\_iter\_seq\_info}, \texttt{sock\_map\_iter\_seq\_info},\\
          \texttt{sock\_hash\_iter\_seq\_info}, \texttt{ksym\_iter\_seq\_info},\\
          \texttt{bpf\_link\_seq\_info}, \texttt{bpf\_map\_seq\_info},\\
          \texttt{sock\_hash\_iter\_seq\_info}, \texttt{sock\_map\_iter\_seq\_info},\\
          \texttt{ipv6\_route\_seq\_info}, \texttt{iter\_seq\_info},\\
          \texttt{ksym\_iter\_seq\_info}, \texttt{netlink\_seq\_info},\\
          \texttt{tcp\_seq\_info}, \texttt{udp\_seq\_info},\\
          \texttt{unix\_seq\_info}, \texttt{bpf\_prog\_seq\_info}\\
      \end{tabular}
  \\ \hline \end{tabular} } \end{table}

	\section*{Appendix B: BPF CVE List}
	\label{app:cve_list}
  We provide the list of evaluated BPF CVEs in \T~\ref{table:cve_table}. 

\begin{table}[H] \caption{Evaluated BPF CVEs.} \label{table:cve_table} 
\centering
  \resizebox{.8\linewidth}{!}{ \begin{tabular}{l} \hline \textbf{CVE ID}
      \\\hline \begin{tabular}[c]{@{}l@{}} 2016-2383, 2017-16995, 2017-16996,
        2017-17852, 2017-17853, 2017-17854,\\ 2017-17855, 2017-17856,
        2017-17857, 2017-17862, 2017-17863, 2017-17864,\\ 2018-18445,
        2020-8835, 2020-27194, 2021-34866, 2021-3489, 2021-3490,\\ 2021-20268,
        2021-3444,2021-33200, 2021-45402, 2022-2785, 2022-23222,\\2023-39191,
        2023-2163
				 \end{tabular}    \\
				\hline         
		\end{tabular}}
	\end{table}

  \section*{Appendix C: Supported BPF Helpers} \label{app:helper} \mr{We list
    the helper functions that are tested on \tool in \T~\ref{table:helper}. To
    test these helper functions, we adapt the BPF programs included in the
    Linux kernel tree to invoke these helpers. We also check their results to
    ensure these helpers are executing correctly on a \tool-enabled system.}

    \begin{table}[htbp] \caption{\tool-supported helpers.} \label{table:helper}
      \centering \resizebox{.8\linewidth}{!}{ \begin{tabular}{cl} \hline
          \textbf{Type} & \textbf{Supported BPF Helpers} \\ \hline
          \multicolumn{1}{c}{\begin{tabular}[c]{@{}c@{}}Map\end{tabular}} &
          \begin{tabular}[l]{@{}l@{}} \texttt{bpf\_map\_lookup\_elem},
            \texttt{bpf\_map\_update\_elem},\\ \texttt{bpf\_map\_delete\_elem},
            \texttt{bpf\_map\_push\_elem},\\ \texttt{bpf\_map\_pop\_elem},
            \texttt{bpf\_map\_peek\_elem} \end{tabular} \\ \hline
          \multicolumn{1}{c}{\begin{tabular}[c]{@{}c@{}}String\end{tabular}} &
          \begin{tabular}[l]{@{}l@{}} \texttt{bpf\_strtol},
            \texttt{bpf\_strtoul}, \texttt{bpf\_strncmp} \end{tabular} \\
          \hline Utilities & \begin{tabular}[l]{@{}l@{}}
            \texttt{bpf\_trace\_vprintk}, \texttt{bpf\_get\_retval},\\
            \texttt{bpf\_set\_retval}, \texttt{bpf\_user\_rnd\_u32},\\
            \texttt{bpf\_get\_raw\_cpu\_id},
            \texttt{bpf\_get\_smp\_processor\_id},\\
            \texttt{bpf\_ktime\_get\_ns}, \texttt{bpf\_ktime\_get\_boot\_ns},\\
            \texttt{bpf\_ktime\_get\_coarse\_ns},
            \texttt{bpf\_get\_current\_pid\_tgid},\\
            \texttt{bpf\_get\_current\_uid\_gid}, \texttt{bpf\_jiffies64},\\
            \texttt{bpf\_get\_attach\_cookie} \end{tabular} \\ \hline Cgroup &
            \begin{tabular}[l]{@{}l@{}} \texttt{bpf\_get\_current\_cgroup\_id},
              \texttt{bpf\_get\_cgroup\_classid\_curr},\\
              \texttt{bpf\_get\_cgroup\_classid},
              \texttt{bpf\_skb\_cgroup\_id},\\
              \texttt{bpf\_sk\_ancestor\_cgroup\_id},
              \texttt{bpf\_skb\_cgroup\_classid},\\
              \texttt{bpf\_sk\_cgroup\_id},
              \texttt{bpf\_skb\_ancestor\_cgroup\_id},\\
              \texttt{bpf\_get\_current\_ancestor\_cgroup\_id} \end{tabular} \\
            \hline Tracing & \begin{tabular}[l]{@{}l@{}}
              \texttt{bpf\_probe\_read\_compat\_str},
              \texttt{bpf\_probe\_read\_compat},\\
              \texttt{bpf\_probe\_read\_kernel\_str},
              \texttt{bpf\_probe\_read\_kernel},\\
              \texttt{bpf\_get\_current\_task},
              \texttt{bpf\_get\_func\_ip\_tracing},\\
              \texttt{bpf\_task\_pt\_regs}, \texttt{bpf\_perf\_event\_read},\\
              \texttt{bpf\_perf\_event\_read\_value},
              \texttt{bpf\_perf\_event\_output},\\
              \texttt{bpf\_get\_func\_ret}, \texttt{bpf\_get\_func\_arg},\\
              \texttt{bpf\_get\_func\_arg\_cnt}, \texttt{bpf\_get\_func\_ip},\\
              \texttt{bpf\_get\_ns\_current\_pid\_tgid} \end{tabular} \\ \hline
              Ringbuf & \begin{tabular}[l]{@{}l@{}}
                \texttt{bpf\_ringbuf\_discard}, \texttt{bpf\_ringbuf\_query},\\
                \texttt{bpf\_ringbuf\_submit},
                \texttt{bpf\_ringbuf\_reserve},\\ \texttt{bpf\_ringbuf\_output}
                \end{tabular} \\ \hline XDP & \begin{tabular}[l]{@{}l@{}}
                \texttt{bpf\_xdp\_fib\_lookup},
                \texttt{bpf\_xdp\_load\_bytes},\\
                \texttt{bpf\_xdp\_store\_bytes},
                \texttt{bpf\_xdp\_adjust\_head},\\
                \texttt{bpf\_xdp\_adjust\_meta},
                \texttt{bpf\_xdp\_adjust\_tail},\\
                \texttt{bpf\_xdp\_get\_buff\_len} \end{tabular} \\ \hline
                Socket & \begin{tabular}[l]{@{}l@{}}
                  \texttt{bpf\_get\_listener\_sock},
                  \texttt{bpf\_skb\_get\_pay\_offset},\\
                  \texttt{bpf\_skc\_to\_mptcp\_sock},
                  \texttt{bpf\_skc\_to\_tcp6\_sock},\\
                  \texttt{bpf\_skc\_to\_tcp\_request\_sock},
                  \texttt{bpf\_skc\_to\_tcp\_sock},\\
                  \texttt{bpf\_skc\_to\_tcp\_timewait\_sock},
                  \texttt{bpf\_skc\_to\_udp6\_sock},\\
                  \texttt{bpf\_skc\_to\_unix\_sock},
                  \texttt{bpf\_sk\_fullsock},\\ \texttt{bpf\_sk\_release},
                  \texttt{bpf\_tcp\_sock},\\
                  \texttt{bpf\_skb\_load\_helper\_8\_no\_cache},\\
                  \texttt{bpf\_skb\_load\_helper\_16\_no\_cache},\\
                  \texttt{bpf\_skb\_load\_helper\_32\_no\_cache},\\
                  \texttt{bpf\_sock\_ops\_cb\_flags\_set},
                  \texttt{bpf\_task\_storage\_delete},\\
              \texttt{bpf\_skb\_load\_bytes},
          \texttt{bpf\_skb\_load\_helper\_16},\\
      \texttt{bpf\_skb\_load\_helper\_32}, \texttt{bpf\_skb\_load\_helper\_8}
  \end{tabular} \\ \hline \\ \end{tabular} } \end{table} 

  \balance
                  \end{document}